\documentclass[aip,graphicx,reprint]{revtex4-1}
\usepackage{graphicx}
\usepackage{amsmath}

\usepackage{cancel}
\usepackage{ulem}
\usepackage{xcolor}

\begin{document}

\title{Thermal circuit model for silicon quantum-dot array structures} 

\author{Takeru Utsugi}
\email[]{takeru.utsugi.qb@hitachi.com}
\affiliation{Research \& Development Group, Hitachi, Ltd., Kokubunji, Tokyo 185-8601, Japan
}
\author{Nobuhiro Kusuno}
\affiliation{Research \& Development Group, Hitachi, Ltd., Kokubunji, Tokyo 185-8601, Japan
}
\author{Takuma Kuno}
\affiliation{Research \& Development Group, Hitachi, Ltd., Kokubunji, Tokyo 185-8601, Japan
}
\author{Noriyuki Lee}
\affiliation{Research \& Development Group, Hitachi, Ltd., Kokubunji, Tokyo 185-8601, Japan
}
\author{Itaru Yanagi}
\affiliation{Research \& Development Group, Hitachi, Ltd., Kokubunji, Tokyo 185-8601, Japan
}
\author{Toshiyuki Mine}
\affiliation{Research \& Development Group, Hitachi, Ltd., Kokubunji, Tokyo 185-8601, Japan
}
\author{Shinichi Saito}
\affiliation{Research \& Development Group, Hitachi, Ltd., Kokubunji, Tokyo 185-8601, Japan
}
\author{Digh Hisamoto}
\affiliation{Research \& Development Group, Hitachi, Ltd., Kokubunji, Tokyo 185-8601, Japan
}
\author{Ryuta Tsuchiya}
\affiliation{Research \& Development Group, Hitachi, Ltd., Kokubunji, Tokyo 185-8601, Japan
}
\author{Hiroyuki Mizuno}
\affiliation{Research \& Development Group, Hitachi, Ltd., Kokubunji, Tokyo 185-8601, Japan
}

\date{\today}

\begin{abstract}
Temperature rise of qubits due to heating is a critical issue in large-scale quantum computers based on quantum-dot (QD) arrays. This leads to shorter coherence times, induced readout errors, and increased charge noise. Here, we propose a simple thermal circuit model to describe the heating effect on silicon QD array structures. Noting that the QD array is a periodic structure, we represent it as a thermal distributed-element circuit, forming a thermal transmission line. We validate this model by measuring the electron temperature in a QD array device using Coulomb blockade thermometry, finding that the model effectively reproduces experimental results. This simple and scalable model can be used to develop the thermal design of large-scale silicon-based quantum computers.
\end{abstract}


\maketitle %\maketitle must follow title, authors, abstract and \pacs

\section{Introduction}
Silicon quantum-dot (QD) arrays are promising candidates for scalable quantum computing platforms because of their outstanding transistor integration, long coherence time, and high fidelity~\cite{Burkard2023,kodera2024trends,Vandersypen2017,Veldhorst2017,Borsoi2022,Li2018,Philips2022,Lee2020,Lee2022}. For realizing a large-scale integration, heating in the QD array is an unavoidable issue. The rise in electron (or hole) temperature due to heating leads to shorter coherence times, induced readout errors, and increased charge noise~\cite{huang2024high,kodera2024trends}. There are several heat sources (e.g., heat and thermal noise inflows through wires, loss of microwave signals, and driving in charge sensors) that are expected to escalate with the large-scale integration of qubits~\cite{Krinner2019,Savin2006,Kawakami2013,takeda2018optimized,Noah2023,undseth2023hotter}.

Two approaches have been investigated to address the heating problem. One is to develop cryo-electronics, effectively reducing the wiring from room temperature and reducing the heat input from the wiring~\cite{Xue2021,pauka2021cryogenic,bartee2024spin}. The other is operating qubits at high temperature using heat-tolerant techniques, such as a Pauli spin blockade readout, which has been demonstrated to operate at temperatures as high as one kelvin~\cite{Petit2018,Ono2019,Petit2020,Yang2020,Petit2022,Camenzind2022,huang2024high}. 
Meanwhile, we are exploring another approach, namely thermal management through the design of heat inflow paths in QD array structures and the control of heat sources using real-time ambient temperature measurements.
To achieve this, the thermal conduction characteristics of the device need to be understood. To accurately measure the thermal characteristics, local heaters and thermometers must be integrated into the QD array.
%The electron temperature rise can be suppressed by increasing the thermal resistance of the heat flow path between the qubit and the heat source and decreasing the thermal resistance between the heat source and the cooler. 

Coulomb blockade thermometry (CBT) is known to measure temperature in QD structures in cryogenic environments~\cite{Beenakker1991,Giazotto2006,rossi2012electron}. CBT has been used to study the thermal conduction properties of quantum point contacts (QPCs)~\cite{Molenkamp1992,Chiatti2006} and single-electron transistors (SETs)~\cite{Hoffmann2007,Hoffmann2009,Dutta2017} in cryogenic environments. However, the thermal conduction characteristics of QD array structures have yet to be investigated. Recently, thermal analysis towards silicon quantum computers has been conducted. Heating in cryo-electronic circuits was investigated using two transistors for the local heater and CBT sensor~\cite{de2023measurement}, and thermal transient in QD array structure was obtained by using microwave pulses and reflectometry~\cite{champain2023real}. Furthermore, the thermal circuit model based on the lumped-element model was developed to analyze the cryogenic CMOS on-chip thermometry, a crucial step for understanding the thermal characteristics in cryo-chips~\cite{Noah2023}.

In this paper, to clarify the thermal conduction characteristics of QD array structures, we propose a simple thermal circuit model. 
In Sec. II, we explain our model, where the QD array structure is described by an effective thermal circuit based on the distributed-element model. This results in an analytical expression of the thermal transmission line. 
In Sec. III, we show the experimental validation of our model, where we measure the electron temperature in a QD array device using CBT. A local heater is implemented by flowing current through the barrier gates of the QD array. This experimental setup mimics the heat generated by the local current flowing through the gate wiring for the qubit addressing~\cite{Kawakami2013,Li2018} and by the microwaves applied to the gate wiring for an electric dipole spin resonance~\cite{Burkard2023}. From the distance dependence between the local heater and the SET, we find that this model reproduces the experimental results. 
In Sec. IV, we discuss the limitations of our model and future work.
Our proposed model is simple, intuitive, and scalable and can be used for the thermal management of large-scale quantum computers in silicon.

\section{Model}
\begin{figure}
\includegraphics[width=80mm]{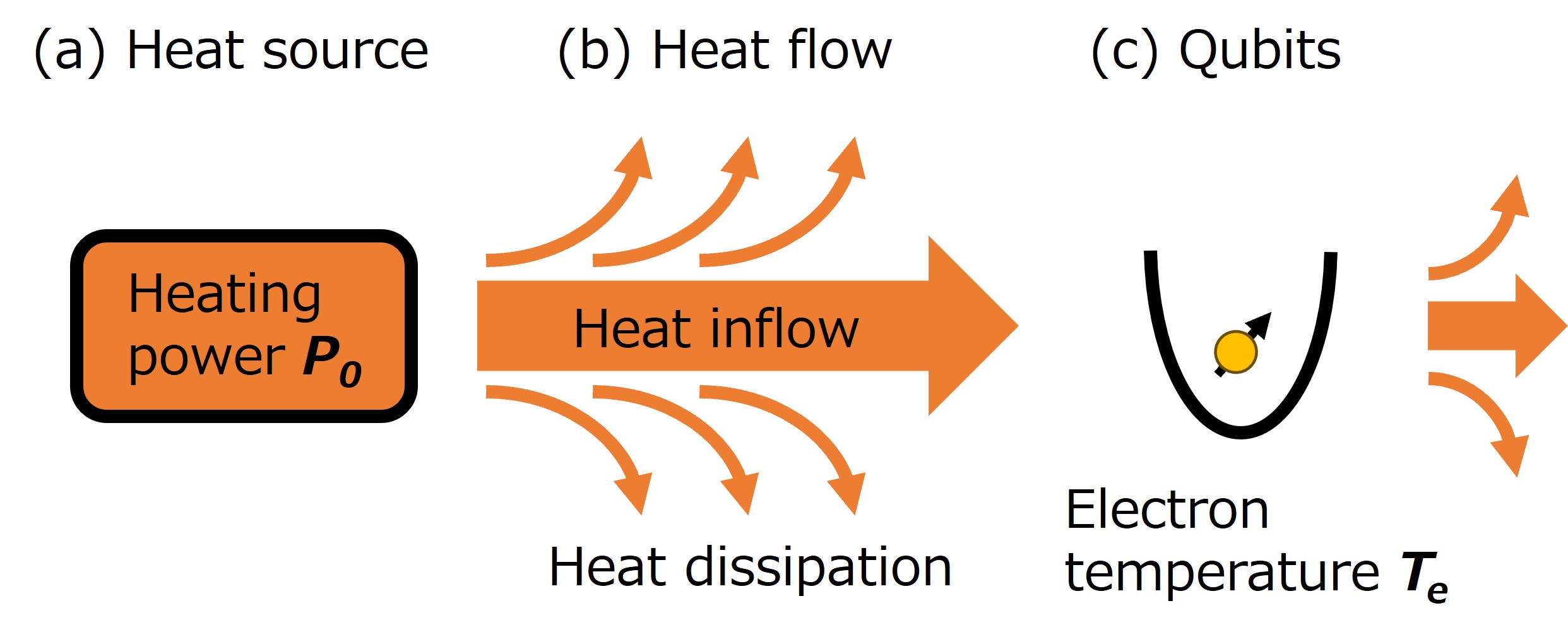}%
\caption{\label{fig0}Conceptual diagram of the heating effects in silicon quantum computers. Heat flow is modeled by describing the electron temperature $T_e$ of qubits as a function of the heating power {$P_0$} and various parameters that depend on the device structure.}
\end{figure}

We describe the heating effects in silicon quantum computers separately in three parts: (a) {Heat} source, (b) heat flow, and (c) qubits. Figure~\ref{fig0} illustrates an conceptual diagram. In silicon quantum computers that are typically implemented in cryogenic environments using dilution refrigerators, there are several heat sources such as radiation within dilution refrigerator, heat and thermal noise inflows through wires, loss of microwave signals, and local currents in charge sensors~\cite{Krinner2019,Savin2006,Kawakami2013,takeda2018optimized,Noah2023,undseth2023hotter}. These heat sources can be modeled as heating power {$P_0$}. The heat inflow to the qubits and the heat dissipation to the cooler depend on the device structure, specifically its thermal conductance and capacitance. Our objective is to model the electron temperature of qubits $T_e$ as a function of {$P_0$} and various parameters depending on the device structure. By employing this model, we can estimate $T_e$ from {$P_0$} to design the heat flow path accordingly.

\begin{figure}
\includegraphics[width=80mm]{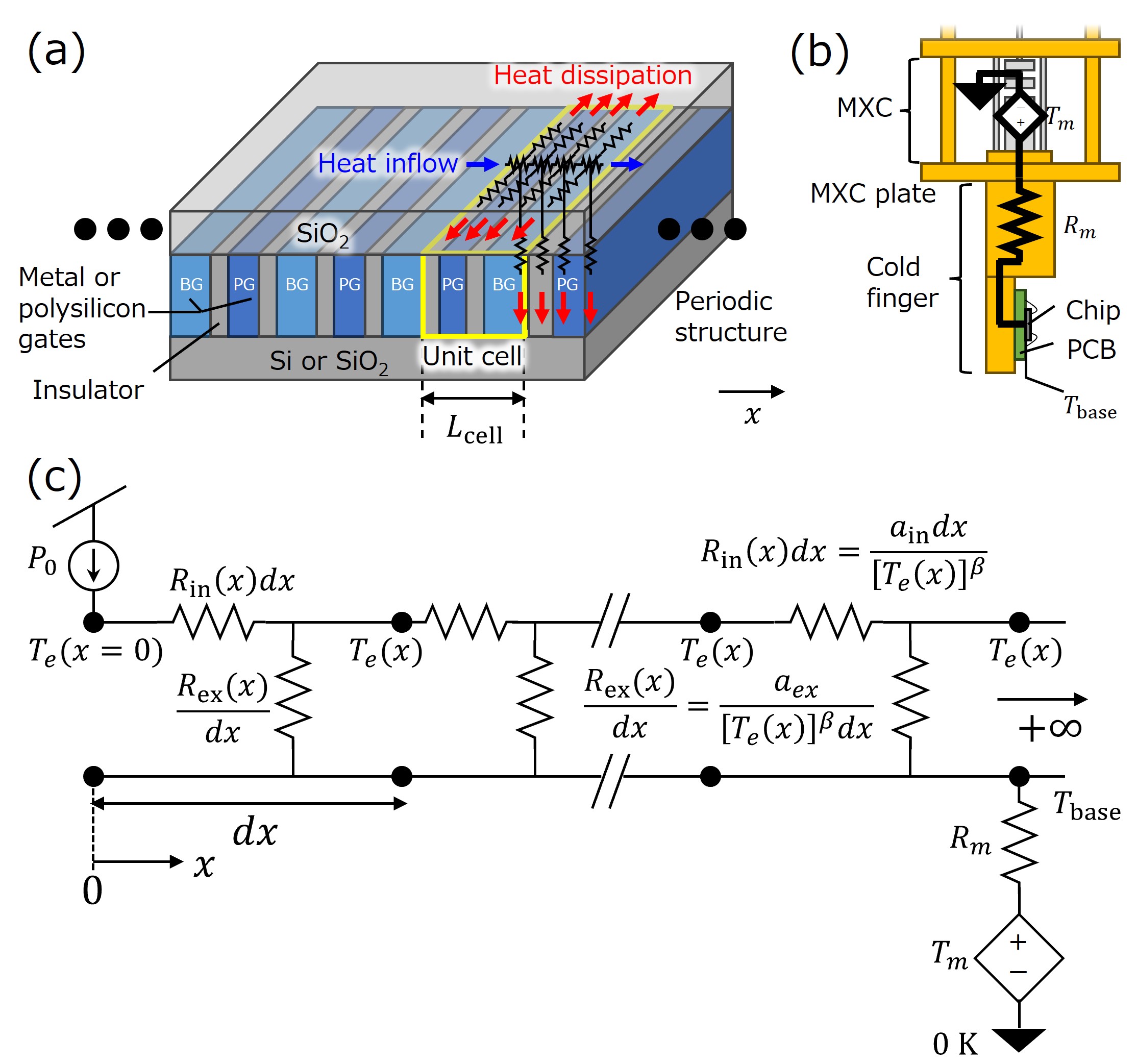}%
\caption{\label{fig1}(a) Model of the silicon QD array structure. The periodic gate structure comprises the plunger gate (PG) and the barrier gate (BG) made of metal or polysilicon. Top and bottom layers are thick Si or SiO$_{2}$. Heat inflow propagates to the $x$ direction, and heat dissipation propagates to the other directions. {This conceptual diagram shows a case where the heat source gate and the measurement point are located at a certain distance from each other.} {(b) Thermal circuit model from the MXC to the chip in the dilution refrigerator. (c) Thermal circuit model for the periodic silicon QD array structure.} }
\end{figure}

For constructing the thermal model of the silicon QD array, we note that typical silicon QD array structures are configured with a periodic gate array~\cite{Vandersypen2017,Veldhorst2017,Borsoi2022,Li2018,Philips2022,Lee2020,Lee2022}, and this periodicity is essential for modeling their thermal characteristics. In our model, the QD array device is represented as a periodic structure shown in Fig.~\ref{fig1}(a). The unit cell comprises a set of a plunger gate (PG), barrier gate (BG), and insulators such as SiO$_{2}$, where the length of the unit cell in $x$ direction is $L_{\rm{cell}}$. For simplicity, the silicon channel and the regions outside the gate structure are omitted. 
{Heat is generated at a gate, and a portion of it propagates toward a qubit located some distance away. All of the heat is eventually dissipated outside the chip. Accordingly, }{t}his heat flow is divided into two paths: heat inflow path and dissipation path. The heat inflow path represents the propagation of heat from the source to the qubits along the periodic gate structure, while the dissipation path represents the heat propagation from the heat source to the cooler. 
Here, the heat is dissipated {mainly} through a medium with {relatively} high thermal conductivity, such as metallic {or polysilicon} gates, {thick SiO$_{\rm{2}}$ or Si at the bottom layer}, metallic wires, and metallic structure in a dilution refrigerator.

We employ a thermal circuit model where thermal characteristics are analogously represented to an electrical circuit. In this model, heating power and temperature are represented by current and voltage, respectively, similar to an electrical circuit model~\cite{Noah2023}.
{The overall dilution refrigerator is modeled to set the base temperature of the QD array device, denoted by $T_{\rm{base}}$, as shown in Fig.~\ref{fig1}(b), based on Ref.~\cite{Noah2023}. Assume that the mixing chamber (MXC) is cooled to a stable temperature $T_m$, where the cooling power and heating power are balanced. The effective temperature-dependent thermal resistance, denoted by $R_m$, includes the thermal resistance of the MXC plate, the cold finger, the printed circuit board (PCB), and wiring from the MXC plate to the chip.}

{The} periodic structures {of the QD array} can be modeled using a thermal distributed-element model, analogous to the transmission line model in electrical circuits. We show the thermal circuit diagram {for the QD array structure} in Fig.~\ref{fig1}({c}), where unlike in the case of electronic circuits, {capacitors and} inductors are excluded. This thermal circuit can be analyzed using {{simultaneous differential equations analogous to}} the telegrapher's equations. In this context, $R_{\rm{in}}$ (heat inflow resistance) {and} $R_{\rm{ex}}$ (heat dissipation resistance) are defined as distributed elements, as shown in Fig.~\ref{fig1}({c}).
{The low side reference of this thermal transmission line is $T_{\rm{base}}$ and the electron temperature rise from $T_{\rm{base}}$ at position $x$ is denoted by $T_{e}(x)$.}

Here, we outline the assumptions underlying our model. (i) The first assumption is that the thermal circuit consists of a {semi-}infinite periodic structure. This assumption is valid for large-scale QD array structures and thermally uniform structures such as those incorporating dummy metals in the gate layer. (ii) The second assumption is that the distributed elements of the thermal circuit, $R_{\rm{in}}$ {and} $R_{\rm{ex}}$ are temperature dependent and spatially {varied}. To account for the temperature dependence of these spatially {non-}uniform circuit elements, we consider the device's {local} effective temperature. (iii) The third assumption is that this {local} effective temperature is proportional to {$T_e(x)$}. These assumptions allow us to use a {modified} transmission line model based on the distributed-element model.

We then explain the thermal circuit elements $R_{\rm{in}}$ {and} $R_{\rm{ex}}$, respectively. $R_{\rm{in}}$ is the thermal resistance (per unit length) for heat inflow from the heat source to the temperature measurement point (qubit). This thermal resistance is a series of resistances of a periodic structure, which is composed of the metallic or polysilicon gates and the insulator (we show the order estimation of $R_{\rm{in}}$, see Supplementary Material II). 
{On the basis of the above discussion, we can write $R_{\rm{in}}(x) =a_{\rm{in}}/(T_e(x)+T_{\rm{base}})^{\beta_{\rm{in}}}$, where $a_{\rm{in}}$ is a constant and $\beta_{\rm{in}}$ depends on the thermal conductivity in cryogenic regime and the thickness of the gate and insulator. Namely, $R_{\rm{in}}(x)$ depends only on $T_e(x)$ due to the local effect. For example, $\beta_{\rm{in}}\approx1$ when the gate is dominant and $\beta_{\rm{in}}\approx2$ when the insulator (amorphous SiO$_{2}$) is dominant (see Supplementary Material I and Refs.~\cite{metal,ekin,aist}).}

$R_{\rm{ex}}$ is the thermal resistance (inverse of the thermal conductance per unit length) for heat dissipation from the heat source. Heat is dissipated through metallic wires and metallic structures in a dilution refrigerator and is finally collected in the 
{MXC}, which acts as a cooler. Therefore, {potentially,} $R_{\rm{ex}}$ depends on the structure of the whole chip and chip’s implementation form such as the circuit board mounting the chip, bonding wires, sample packaging, and cables. 
{However, we assume that the temperature dependence of this resistance is mainly determined by the local effective temperature, as mentioned in the assumption (ii). This is because, along the heat propagation path, regions near the gate structure have high thermal resistance, while regions farther away—such as metal wiring layers, the bottom layer, and metallic structures—have relatively low thermal resistance. As a result, large thermal gradients are likely to form near the gate structure, whereas the more distant regions remain close to $T_{\rm{base}}$. This assumption holds when a good heat dissipation structure is present around the gate region.} 
{We then write $R_{\rm{ex}}(x) = a_{\rm{ex}}/(T_e(x)+T_{\rm{base}})^{\beta_{\rm{ex}}}$, where $\beta_{\rm{ex}}=$1--2, considering the composition of the local region.}

%%%%%%%%%%%%%%%%%%%%%%%%%%%%%%%%%%%%%
{We formulate the relations between the electron temperature $T_e(x)$ and the heat flow $P(x)$ as following simultaneous differential equations:
\begin{align}
-\frac{dT_e(x)}{dx}&=R_{\rm{in}}(x)P(x),\label{EQ001}\\
-\frac{dP(x)}{dx}&=\frac{1}{R_{\rm{ex}}(x)}T_e(x).\label{EQ002}
\end{align}
Here, we assume that $T_e(x) \gg T_{\rm{base}}$ and $\beta_{\rm{in}} \approx \beta_{\rm{ex}} \, (\equiv \beta)$, which are referred to as assumptions (iv) and (v), respectively.
Assumption (iv) is valid when thermal effects are significant—precisely the regime in which this analysis becomes relevant.
The validity of assumption (v) is supported when the temperature dependence of the gate and insulator is similar, or when one of the ratios dominates, both in the local region (as considered in assumption (ii)).
On the basis of above discussion, we can reasonably set $\beta \approx 1\text{--}2$.}
%Furthermore, when the insulator thickness is relatively small compared to that of the gate ($\beta_{\rm{in}} \approx 1$), and considering that metallic materials dominate the dissipation path ($\beta_{\rm{ex}} \approx 1$)---as discussed in Supplementary Material I---this supports the approximation $\beta \approx 1$. 

{Based on these assumptions, we can rewrite the equations into an analytically solvable form as follows:\begin{align}
-\frac{dT_e(x)}{dx}&=\frac{a_{\rm{in}}}{\left[T_e(x)\right]^\beta}P(x),\label{EQ003}\\
-\frac{dP(x)}{dx}&=\frac{\left[T_e(x)\right]^{\beta+1}}{a_{\rm{ex}}}.\label{EQ004}
\end{align}
Considering boundary conditions of $P(x=0)=P_0\geq0$ and $P(x=\infty)=T_e(x=\infty)=0$,
we can obtain one of the solutions for $x\geq0$ and $P(x)\geq0$ as
\begin{align}
T_e(x)&=\left[a_{\rm{in}}a_{\rm{ex}}(\beta+1)\right]^{\frac{1}{2(\beta+1)}}e^{-\sqrt{\frac{a_{\rm{in}}}{a_{\rm{ex}}(\beta+1)}}x}P_0^{\frac{1}{\beta+1}}\notag\\
        &=a e^{-\frac{x}{L_{\rm{th}}}}P_0^{\frac{1}{\beta+1}},\label{EQ005}\\
P(x)&=P_0e^{-\sqrt{\frac{a_{\rm{in}}(\beta+1)}{a_{\rm{ex}}}}x}.\label{EQ006}
\end{align}
where a coefficient $a\equiv[a_{\rm{in}}a_{\rm{ex}}(\beta+1)]^{\frac{1}{2(\beta+1)}}$ and a thermal characteristic length $L_{\rm{th}}\equiv\sqrt{\frac{a_{\rm{ex}}(\beta+1)}{a_{\rm{in}}}}$ are introduced.}
%%%%%%%%%%%%%%%%%%%%%%%%%%%%%%%%%%%%%
The main result in this section, Eq.~(\ref{EQ005}), indicates that the electron temperature is modeled as a function of heating power {$P_0$}, the distance {$x$}, and three device-dependent parameters {($a$, $L_{\rm{th}}$, and $\beta$).}

{We note that the parameter $a$ depends on the multiple of the thermal resistances coefficients ($a_{\rm{in}}\times a_{\rm{ex}}$). On the other hand, $L_{\rm{th}}$ depends on the ratio of the thermal resistances coefficients ($a_{\rm{ex}}/a_{\rm{in}}$). Furthermore, both $a$ and $L_{\rm{th}}$ are independent of $T_e$. The term $L_{\rm{th}}$ represents the characteristic length of the heat dissipation, indicating that the heating effect decreases by $1/e$. Thereby, the temperature independent parameters $a$ and $L_{\rm{th}}$ thermally characterize the QD array structure.}

{
Here, background heating power is introduced to fit the experimental results using the proposed model, Eq.~(\ref{EQ005}). The total electron temperature rise, denoted by $T_e^{\rm{total}}$, is described as
}
\begin{equation}
{
T_e^{\rm{total}} = \left(T_e^{\beta+1} + T_B^{\beta+1}\right)^{\frac{1}{\beta+1}} = a e^{-\frac{x}{L_{\rm{th}}}} \left(P_0 + P_B\right)^{\frac{1}{\beta+1}},
}
\end{equation}
{
where $T_B = a e^{-\frac{x}{L_{\rm{th}}}} P_B^{\frac{1}{\beta+1}}$, and $P_B$ represents the background heating power.
}
We use this equation to fit the experimental data. Note that in our experiment, {$T_B$ and $P_B$} include the effects of lifetime broadening and charge noise broadening~\cite{de2023measurement}, {whereas $P_0\gg P_B$ typically holds in most cases.}

{
Finally, the validity of the model is confirmed for a finite and discrete structure. Although assumption (i) considers a semi-infinite structure, the proposed model, Eq.~(\ref{EQ005}), also accommodates similar circuit configurations for finite systems. The detailed calculation method is provided in Supplementary Material III. These results indicate that the device structure used in the experiment remains valid, even if it deviates from the semi-infinite configuration assumed in assumption (i).
}

%############################################
\section{Experiment}
\subsection{Device structure}
To verify the proposed thermal model, we experimentally measure $T_e$ in a QD array. Figure~\ref{fig2}(a) shows the device structure~\cite{utsugi2023single}. In this device, we can measure the temperature at different distances from the heating point. The T-shaped {silicon-on-insulator (SOI)} channel (green) is fabricated, and multiple polysilicon gate electrodes are formed on the T-shaped Si channel. The horizontal channel is covered with first gates (BG0-BG3, light blue) and second gates (PG0-PG3, blue), for a total of seven gates. These fine-pitch gate structures are fabricated using the self-align patterning process~\cite{Lee2020,kuno2024}. Each gate has terminals on both sides, allowing current to flow through them by applying DC voltages through wiring connected to instruments at room temperature.

Another second gate (SGS, red) and other third gates (TG1 and TG2, orange) on the vertical channel comprise a single-electron transistor (SET), and temperature around the SET can be measured using CBT as described below. In this experiment, we treat the temperature measured by the SET as the electron temperature $T_e$. 

The four first gates (BG0-BG3) are used as heat sources by flowing current through each gate. Figure~\ref{fig2}(b) shows the cross-section between A and A’ in Fig.~\ref{fig2}(a). Corresponding to the model shown in Fig.~\ref{fig1}(a), the unit cell length of this device is $L_{\rm{cell}}=120$ nm (we also measure the device of $L_{\rm{cell}}=160$ nm).

The silicon chip with the above structure is glued to a cryogenic printed circuit board (QBoard, Qdevil) using silver paste, and each terminal is wired using aluminum bonding wire. Measurements are performed using a dilution refrigerator (Proteox, Oxford Instruments) with a base temperature of 8 mK.

\subsection{Measurement}
We use CBT to estimate the local temperature at the SET. In CBT, there are several restrictions for QD parameters for accurately measuring temperature, e.g., the tunnel rate of barriers of SET $h\Gamma$, energy spacing of the quantum levels $\Delta E$, thermal energy $k_BT$, and charging energy $e^2/C$~~\cite{Beenakker1991}. 
In this experiment, we use the classical regime, i.e., $h\Gamma \ll \Delta E \ll k_BT \ll e^2/C$. 
{This condition is justified as follows: 
$h\Gamma$ is less than $0.03~\mathrm{meV}$ when the current ($I_d=e\Gamma$) is below $1~\mathrm{nA}$, 
$\Delta E$ is estimated to be less than $0.1~\mathrm{meV}$ based on the device structure~\cite{kouwenhoven1997electron}, 
$T_e$ in our measurements ranges from $1$ to $30~\mathrm{K}$, corresponding to $k_B T_e = 0.086$--$2.6~\mathrm{meV}$, 
while $e^2/C$ is about $5$--$10~\mathrm{meV}$ from the measured Coulomb diamond.}
{The experimental setup is also verified by the measurement of MXC temperature and $T_e$ (see Supplementary Material IV).}

By measuring the current $I_d$ flowing through the SET while sweeping the voltage of the SGS gate $V_g$ and the drain $V_d$ [Fig.~\ref{fig3}(a)], we can obtain the SGS gate voltage dependence of the differential conductance $G=d I_d/dV_d$ [Fig.~\ref{fig3}(b)] and estimate the electron temperature around SET as $T_e$.
{Note that the Coulomb diamonds shown in Fig.~\ref{fig3}(a) and \ref{fig3}(b) deviate from the ideal shape, indicating the possible formation of multiple QDs near or within the SET. This is likely due to structural imperfections in the device, such as disorder or surface roughness in the SET channel. To mitigate the influence of such imperfections, we select a Coulomb diamond that is as well-isolated as possible.
Furthermore, the bias offset in the source-drain voltage $V_{\rm{ds}}$ observed in Fig.~\ref{fig3}(a) and \ref{fig3}(b) originates from imperfections in the measurement system—specifically, an offset in the current measurement. This may be attributed to the Seebeck effect in the cables or fluctuations in the ground level in the source measure unit. However, this offset does not significantly affect the measurement results, as it is appropriately corrected by $V_{\rm{correct}}$ during data analysis, i.e., $V_d=V_{\rm{sd}}-V_{\rm{correct}}$.}

The differential conductance $G$ in the region where $V_d$ is small is~\cite{Beenakker1991}
\begin{equation}
\frac{G}{G_{\rm{max}}}=\frac{\alpha e (V_g-V_0)}{k_B T_e \sinh\left(\frac{\alpha e(V_g-V_0)}{k_B T_e}\right)} \approx \cosh^{-2}\left(\frac{\alpha e (V_g-V_0)}{2.5k_B T_e}\right),\label{Gform}
\end{equation}
where $G_{\rm{max}}$, $\alpha$, $e$, $k_B$, and $V_0$ are the peak value of $G$, the lever arm, the elementary charge, the Boltzmann constant, and the voltage value at $G_{\rm{max}}$, respectively. The lever arm was estimated as $\alpha = 0.08$ by obtaining the Coulomb diamond shown in Fig.~\ref{fig3}(a). Figure~\ref{fig3}(c) and \ref{fig3}(d) show examples of $G$ (Coulomb peak) at different $T_e$ situations. It can be confirmed that the Coulomb peak shape reflects the change in $T_e$, and $T_e$ can be estimated by fitting the experimental result by the curve of Eq.~(\ref{Gform}). {This measurement method is constrained by the tunnel rate and is applicable in the regime where $k_B T_e \gg h\Gamma$. Given that $h\Gamma/k_B < 0.3~\mathrm{K}$ in this experiment, the condition $k_B T_e \gg h\Gamma$ implies that the method is applicable for $T_e \gtrsim 1~\mathrm{K}$.}
\begin{figure}
\includegraphics[width=80mm]{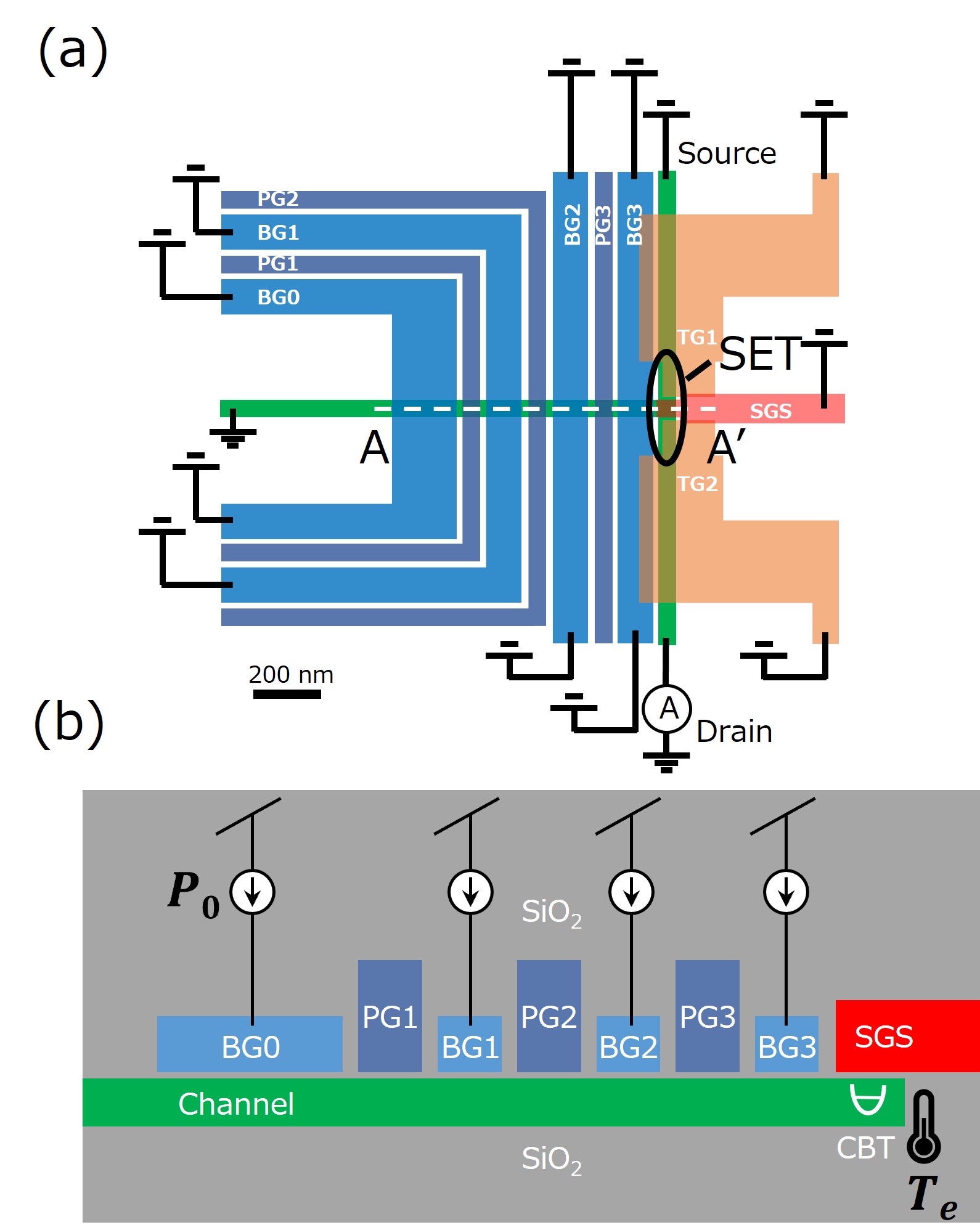}%
\caption{\label{fig2}(a) Schematic diagram of the device structure and setup for electron temperature measurement. An SOI channel (green), BGs (light blue), TGs (orange), and SGS (red) have electrodes to apply voltages, where each wiring is a twisted cable made of phosphor bronze and has a low pass filter mounted on Qboard. (b) Schematic diagram of the cross-section at A-A' in (a). The BGs made of polysilicon are used as a heater by flowing current. Gate SGS is used as the plunger gate in SET.}
\end{figure}

We use a local heater by applying a current to each of the four gate electrodes BG0-BG3 as shown in Fig.~\ref{fig4}(a). If the electric field around the SET changes during the measurement, the SET conditions are affected by the electric field and the temperature measurement cannot be performed correctly. Therefore, we apply a voltage of $+V/2$ and $-V/2$, where {$V=\sqrt{RP_0}$} ($R$ is the resistance of the entire gate wiring) to each terminal of the gate to allow current to flow through the gate. Due to the symmetry of the wiring structure in the dilution refrigerator, we set the voltage to be approximately zero on the channel. This stabilizes the SET condition because the voltage condition around the SET is stable even if the heating power {$P_0$} is changed by varying $V$, where {$P_0=V^2/R$}. We measure the resistance of each gate wiring excluding the additional resistance on the circuit board for a low pass filter and use these values to calculate {$P_0$}, considering that the polysilicon gate has the dominant resistance among all of the gate wiring.
\begin{figure}
\includegraphics[width=80mm]{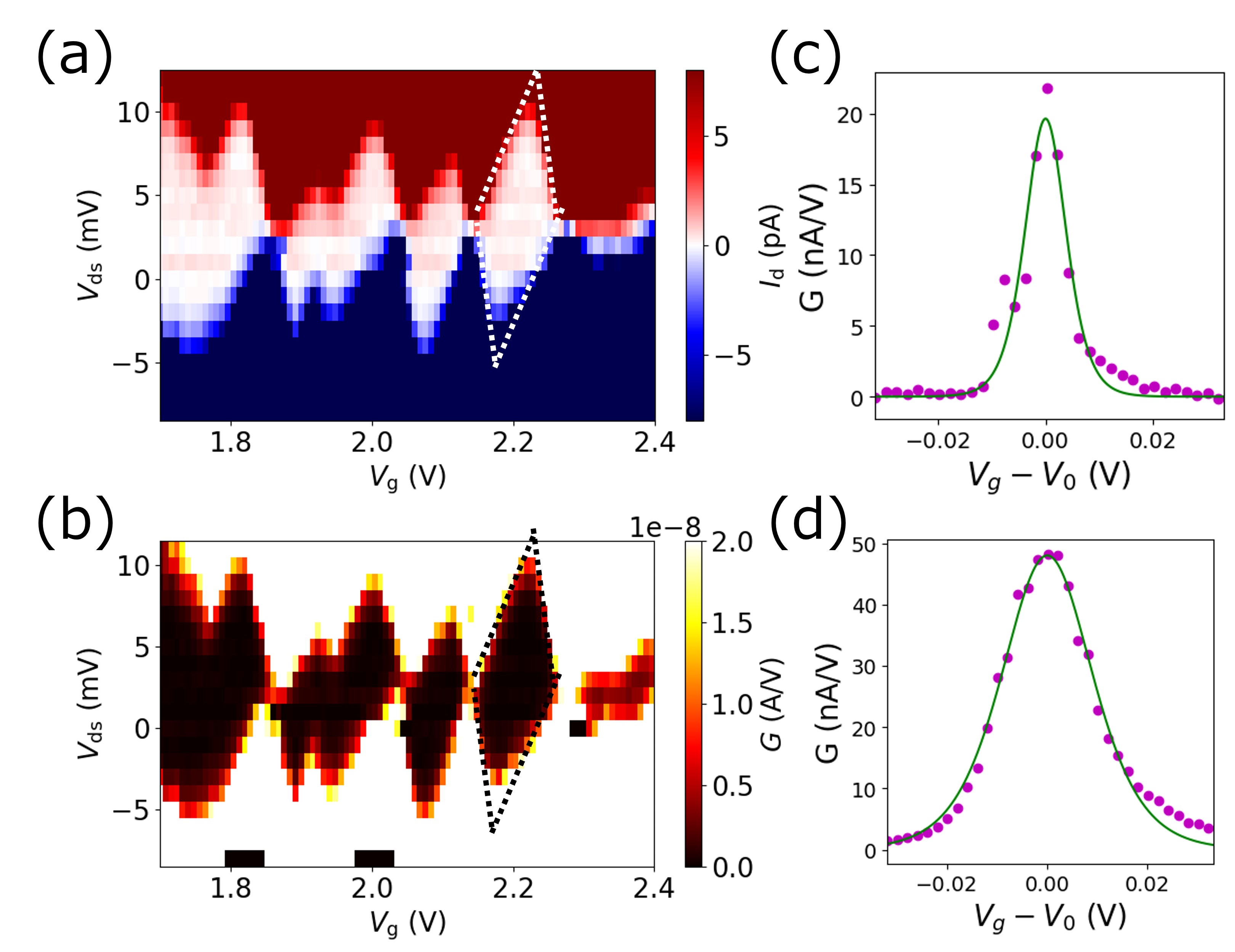}%
\caption{\label{fig3} (a,b) Coulomb diamond of (a) the drain current of SET $I_d$ and (b) the differential conductance of SET $G$. The (a) white and (b) black dotted lines show one of the Coulomb diamond areas. (c) and (d) show examples of Coulomb peak and fitting curve with Eq.~(\ref{Gform}) at different temperatures. The estimated temperatures are (c) 2.4 {$\pm~0.1$}~K and (d) 5.5 {$\pm~0.1$}~K, { where the estimated error is the standard deviation of the $T_e$ estimate. It should be noted that the increase in the conductance peak value in (c) and (d) is most likely attributed to temperature-induced changes in the tunneling rate of the SET formed by the multiple QDs. The bias offset of source-drain voltage $V_{\rm{ds}}$ in (a) and (b) originates from
imperfections in the measurement system, as described in the main text. This offset is appropriately corrected in the estimation of $T_e$.}}
\end{figure}

Figure~\ref{fig4}(b) shows the experimental results of $T_e$ as a function of the heating power {denoted by $P_0$ and the distance between the heater and the SET denoted by $D$} when each gate is used as a heat source. As expected, $T_e$ increases monotonically with {$P_0$} and $1/D$. The experimental results can be fitted using {Eq.~(\ref{EQ005})} as
\begin{equation}
{T_e=A_i (P_0+P_B)^{\frac{1}{\beta+1}}, \label{Te}}
\end{equation}
where $(i\in\{$BG0,BG1,BG2,BG3$\})$ [see the dotted curves in Fig.~\ref{fig4}(b)]. 
The inset of Fig.~\ref{fig4}(b) plots $A_i$ and shows the best-fit curve with {$A_i=a e^{-\frac{D}{L_{\rm{th}}}}$} from Eq.~(\ref{EQ005}).
We experimentally confirm that $T_e$ rises is proportional to $P_0^{\frac{1}{\beta+1}}$ as in the model, and the heating effect of distance is reduced exponentially as expected.
{However, we note that it is difficult to determine from this experiment whether the fitted curve strictly follows exponential decay. To clarify this point, it would be necessary to either increase the number of measurement conditions or use a setup in which the value of $\beta$ is known—i.e., independently measured in advance by an appropriate method.}
We also measure another device of different gate pitch as shown in {Fig.~\ref{fig4}(c)}. From these results, we conclude that the proposed model successfully reproduces the experimental results. 
This demonstrates its capability to model the thermal conduction characteristics of silicon QD array structures effectively.

\begin{figure}
\includegraphics[width=80mm]{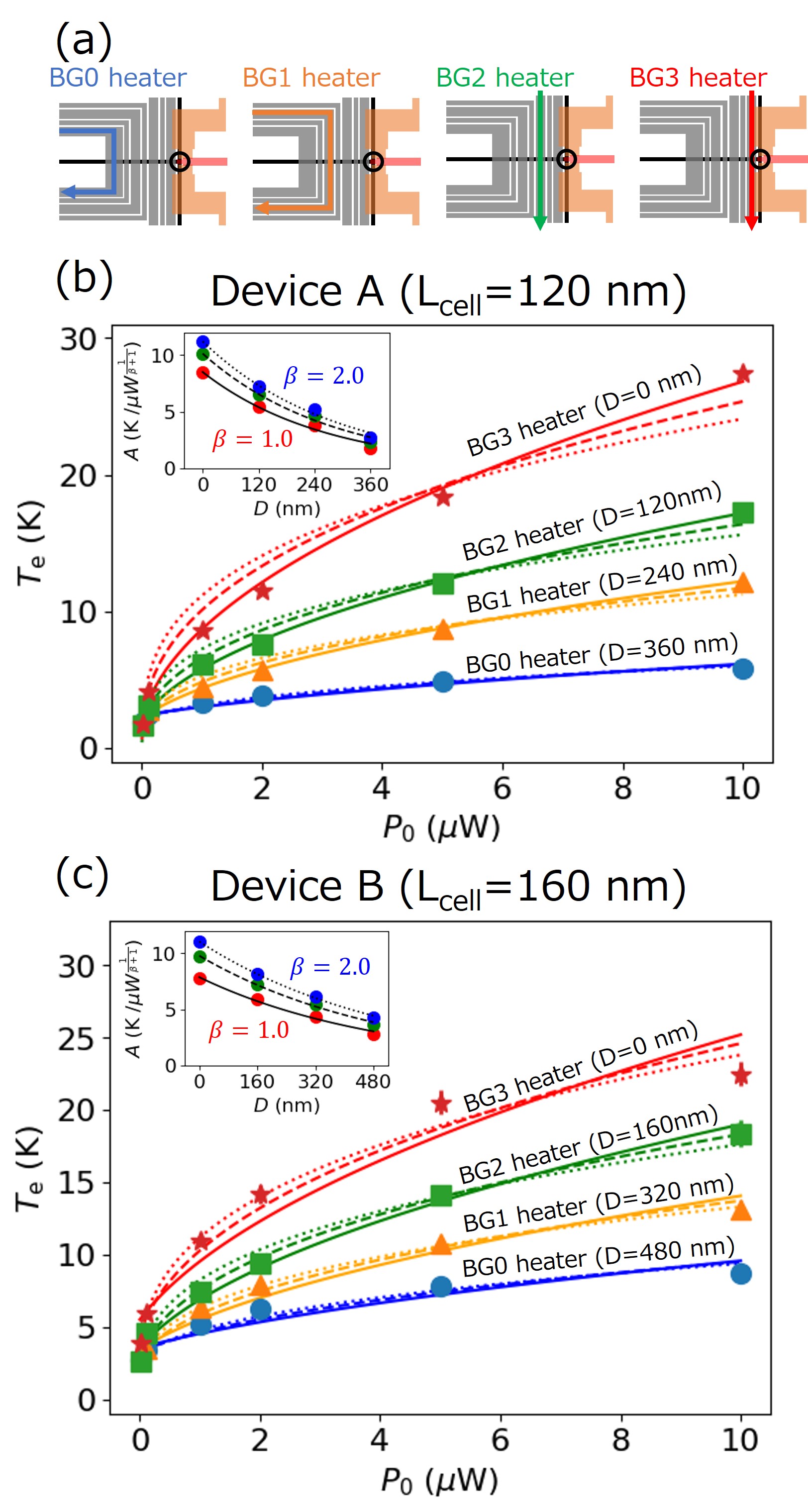}%
\caption{\label{fig4}
(a) Schematic illustration of local heaters formed by applying current to the barrier gates. The resistance of each gate at cryogenic temperatures, excluding the resistance of the low-pass filter on the Qboard, is $R_{\rm{BG0}} = 78.8~\rm{k}\Omega$, $R_{\rm{BG1}} = 78.1~\rm{k}\Omega$, $R_{\rm{BG2}} = 70.9~\rm{k}\Omega$, and $R_{\rm{BG3}} = 61.3~\rm{k}\Omega$, respectively.  
{(b) Experimental results of $T_e$ as a function of {$P_0$} and $D$ in device A with $L_{\rm{cell}} = 120$ nm. The solid, dashed, and dotted curves represent the best-fit results using Eq.~(\ref{Te}) with $\beta = 1.0$, $1.5$, and $2.0$, respectively. The error bars indicate the standard deviation of the estimated $T_e$, although they are smaller than the marker size in the plot (see Supplementary Material IV). The inset shows the best-fit values of $A$ as a function of $D$, with the solid, dashed, and dotted curves corresponding to fits using $A = a e^{-D/L_{\rm{th}}}$ for $\beta = 1.0$, $1.5$, and $2.0$, respectively. The second fit parameter $P_{B}$ ranges from 0 to 2 $\mu$W across all distances, with no clear distance dependence observed. (c) Similar results to (b) obtained from device B with $L_{\rm{cell}} = 160$ nm.
}}
\end{figure}

\section{Discussion}
To consider the limitations of our model, we re-summarize the assumptions for deriving Eq.~(\ref{EQ005}) according to the above discussions: 
(i) The thermal circuit of the model is a {semi-}infinite periodic structure.
(ii) Both $R_{\rm{in}}(x)$ and $R_{\rm{ex}}(x)$ are temperature-dependent and spatially {non-}uniform,
(iii) The {local} effective temperature of the QD array is proportional to $T_e(x)$.
{
(iv) $T_e(x) \gg T_{\rm{base}}$.
(v) $\beta_{\rm{in}} \approx \beta_{\rm{ex}} \, (\equiv \beta)$.
Regarding assumption (i), numerical simulations show that even a finite and discrete circuit exhibits the same dependence as Eq.~(\ref{EQ005}) (see Supplementary Material III).
}
Despite these potential sources of modeling error, the consistency between the experimental results and the model supports the validity of our simplifications.

\begin{table}
\centering
\begin{tabular}{lcc}
\hline\hline
 & Device A & Device B \\
\hline
$L_{\rm{cell}}$ (nm)& 120 & 160 \\
$a$ (K$/\sqrt{\rm{\mu W}}$) & 8.5 & 7.9 \\
$L_{\rm{th}}$ (nm) & 264 & 500 \\
$R_{\rm{in}}T_e$ ($\rm{K^2/(m\cdot W)}$) ~~~~& $2.7\times10^{14}$ & $1.2\times10^{14}$ \\
$R_{\rm{ex}}T_e$ ($\rm{K^2\cdot m/W}$)& $9.5$ & $15.6$ \\
$R^{\rm{est}}_{\rm{in}}$ ($\rm{K^2/(m\cdot W)}$) &$4.0\times10^{14}$&$3.5\times10^{14}$\\
\hline\hline
\end{tabular}
\caption{The device-dependent parameters calculated from the experimental results assuming $\beta=1$, and order estimation results at $T_e=1$ K (the last row).}
\label{table}
\end{table}

{Table~\ref{table} presents the device-dependent parameters estimated from the experiments, where we compare devices A and B. 
To evaluate the thermal resistances, we use the relations of $R_{\rm{in}}T_e^{\beta}=a^{\beta+1}/L_{\rm{th}}$ and $R_{\rm{ex}}T_e^{\beta}=a^{\beta+1}L_{\rm{th}}/(\beta+1)$ derived from the definitions of the $a$ and $L_{\rm{th}}$. Here we use $\beta=1$. We analyze these results to the order estimation values of heat inflow resistance $R^{\rm{est}}_{\rm{in}}$ on the basis of the device structures and the thermophysical properties (see Supplementary Material II). We note that the order estimation of $R_{\rm{ex}}$ is difficult due to the complicated heat dissipation path. We have successfully quantified the thermal characteristics of two QD array structures with different gate pitches. The results indicate that the two devices differ significantly in the characteristic length $L_{\rm{th}}$, while showing only minor variation in the parameter $a$. We believe that further experimental investigations to clarify the origin of this discrepancy will contribute to the optimization of the device structure.}

We also discuss how to design the QD array structure. To mitigate the thermal effects on the qubits, based on Eq.~(\ref{EQ005}), the following conditions should be met: (I) {$P_0$} should be reduced, (II) $D$ should be increased, and (III) the device-dependent parameters $a$ and $L_{\rm{th}}$ should be minimized. Conditions (I) and (II) are straightforward, i.e., the amount of heating power needs to be reduced, for example, by lowering the electrical resistance of the gate, and the qubits need to be kept appropriately distant from the heat sources. Our main result is condition (III), which is useful for quantitative thermal design of the QD array structures. Reducing $a$ implies that even if the heating power is high, the temperature rise is low; in other words, the structure does not easily accumulate heat. 
To reduce $a \propto \sqrt[2(\beta+1)]{R_{\rm{in}} R_{\rm{ex}}}$, the thermal resistance of the entire structure should be minimized. On the other hand, reducing $L_{\rm{th}}$ means creating a structure that makes it difficult to transport heat to the qubits 
{—i.e., thermally separating the heat source from the qubits.}
To reduce $L_{\rm{th}} \propto \sqrt{R_{\rm{ex}}/R_{\rm{in}}}$, the ratio of heat dissipation to heat inflow should be increased. For example, an additional metallic heat sink at the top and bottom of the gate layer can be effective. This can be achieved using technologies such as through-silicon vias (TSVs)~\cite{taguchi2024si}. 

The next challenge is to find a way to simultaneously achieve both high QD (or qubit) performance and thermal tolerance. To further comprehensively optimize the QD array structure, the effect of $T_e$ on the fidelity of quantum computing needs to be quantitatively estimated. The rise in $T_e$ is known to lead to shorter coherence times, degraded readout fidelity, and increased charge noise (low operation fidelity)~\cite{huang2024high,kodera2024trends}. However, these quantitative evaluations have yet to be clarified, except for certain readout fidelity~\cite{keith}, suggesting that further experimental and theoretical studies are necessary.

In future work, we will investigate the thermal dynamic model of the QD array structure to control the heat sources using real-time ambient temperature measurements.
The thermal dynamic model is also strongly related to the operation fidelity of qubits, including the heat-induced frequency shift~\cite{takeda2018optimized,undseth2023hotter,sato2024simulation}. This will be modeled by introducing the heat capacitance in the thermal circuit and measured by a high-speed readout method such as reflectometry~\cite{champain2023real}.
{The simplest case of such a dynamic thermal effect is the heating caused by gate pulsing. In this case, the introduction of heat capacitance leads to a thermal time constant, defined as the product of local thermal resistance and local heat capacitance on the thermal circuit, which governs the transient thermal response. If the time constant is shorter than the pulse duration, the thermal behavior is expected to resemble the static case, and the thermal coupling can still be characterized by $L_{\rm{th}}$. Conversely, if the time constant is longer than the pulse duration, the thermal impact is expected to be reduced. This behavior is analogous to a low-pass filter response in electrical circuits.}

{Another compelling topic is phonon engineering~\cite{maire2017heat}. If we can design a device structure in which phonons propagate coherently as waves, it may be possible to achieve more sophisticated thermal control. We intend to further investigate this concept and consider how our proposed model could be situated within the broader context of phonon engineering.}

In conclusion, we proposed a simple thermal circuit model for the silicon QD array structure and validated our model in experiments. Our proposed model is intuitive, simple, and scalable and is applicable to the wide spread of the QD array structures for thermal analysis.

\section*{Supplementary Material}
The Supplementary Material provides additional information to support the discussion in the main text.

\begin{acknowledgments}
We acknowledge N. Mertig and A. Ramsay at Hitachi Cambridge Laboratory for helpful discussions. This work is supported by JST Mooshot R\&D Grant No. JPMJMS2065.
\end{acknowledgments}

\section*{AUTHOR DECLARATIONS}
\subsection*{Conflict of interest}
The authors have no conflicts to disclose.

\section*{Data Availability}
The data that support the findings of this study are available 
from the corresponding author upon reasonable request.

%%%%%%%%%%%%%%%%%%%%%%%%%%%%%%%%%%%%%%%%%%%%%%%%%%%%%%%%%%%%%%%%
\appendix

\section{Thermophysical properties}
\label{app4}
We summarize thermophysical properties of represented materials in the typical silicon fabrication process based on Refs.~\cite{metal,ekin,aist}, as shown in Supplementary Fig.~\ref{fig8}.
\begin{figure}
\includegraphics[width=70mm]{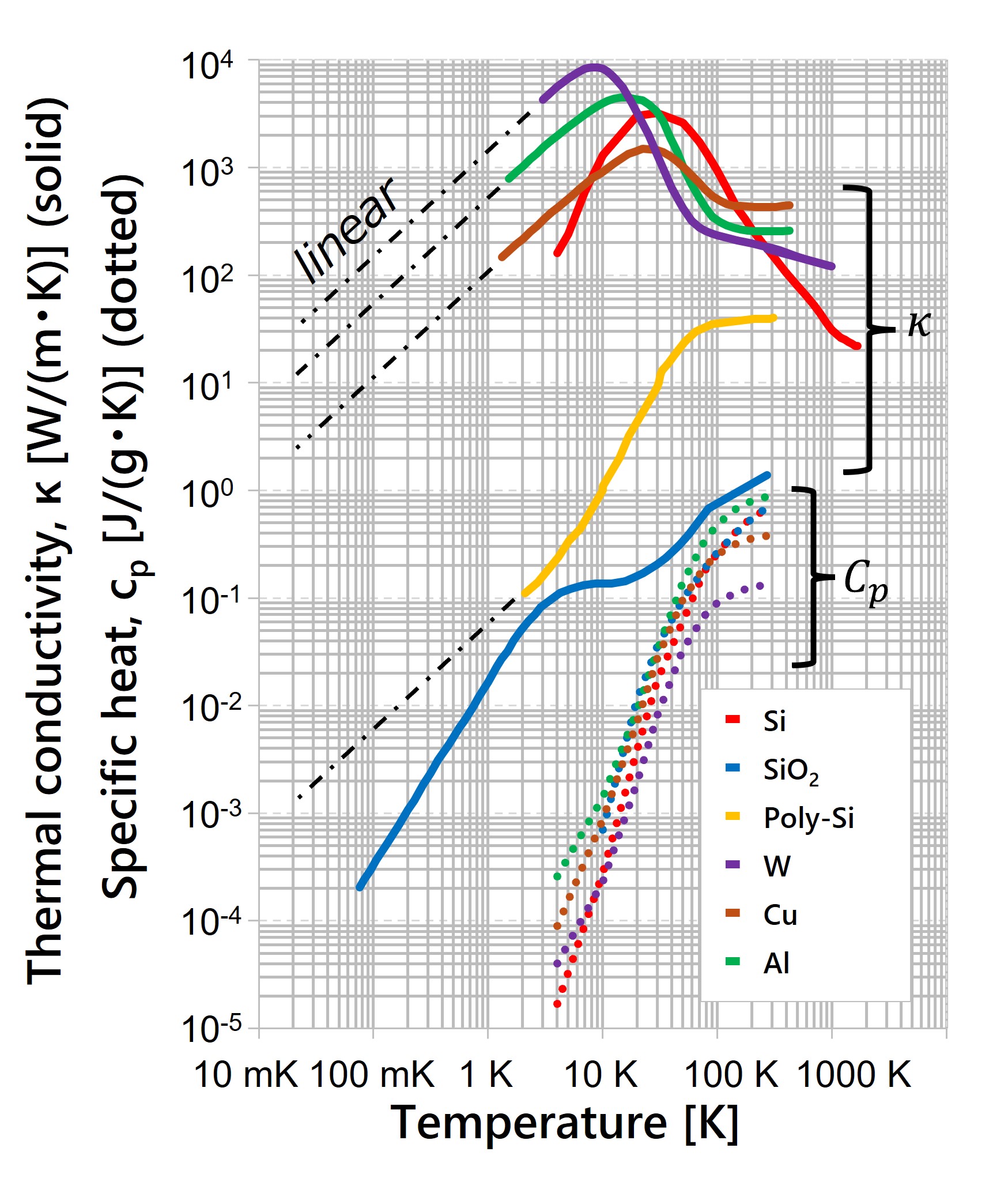}
\caption{\label{fig8}Thermophysical properties of represented materials where $\kappa$ of metallic materials (W, Cu, Al, and Poly-Si) are extrapolated by linear function ($\propto T$) in the cryo-temperature regime~\cite{metal,ekin}. Note that Si is single crystal~\cite{aist}, SiO$_{2}$ is amorphous~\cite{ekin}, and Poly-Si is P-type polysilicon (B-doped = $3\times 10^{20}$ atom/cm$^3$)~\cite{aist}.}
\end{figure}

%\section{Measurement result of device B}
%\label{app3}
%We measure another device with $L_{\rm{cell}} = 160$ nm (device B), similar to the device described in the main text (device A). Supplementary Figure~\ref{fig7} shows the results, which exhibit the same trend as those in Fig.~5 in the main text. However, the electron temperature is lower than that of device A. This is attributed to the larger gate pitch, which implies a longer gate length and a higher ratio of the polysilicon gate to the insulator. This configuration likely decreases the resistance $R_{\rm{in}}$ while maintaining $R_{\rm{ex}}$, resulting in an increased thermal characteristic length $2/\gamma_{\rm{th}} \approx 2\sqrt{R_{\rm{ex}}/R_{\rm{in}}}$. On the other hand, $a \propto \sqrt[4]{R_{\rm{in}}R_{\rm{ex}}}$ changes minimally due to the fourth root. This measurement further validates the proposed thermal model. In Supplementary Material III, we further summarize the experimental results.
%\begin{figure}
%\includegraphics[width=70mm]{007.jpg}%
%\caption{\label{fig7}Experimental results of $T_e$ as a function of $P$ and $D$ in device B of $L_{\rm{cell}}=160$ nm. The dotted curves are the best-fit results with Eq.~(9) in the main text. The inset shows the best-fit results of $A$ as a function of $D$ and the dotted curve is the best-fit result with $A=a e^{-\frac{\gamma_{\rm{th}}}{2} D}$, where $a= 7.0~\rm{K}/\sqrt{\rm{\mu W}}$ and $2/\gamma_{\rm{th}} = 700~\rm{nm}$. }
%\end{figure}

\section{Order estimation of heat inflow resistance}
\label{app5}
%\begin{table}
%\centering
%\begin{tabular}{lcc}
%\hline\hline
% & Device A & Device B \\
%\hline
%$L_{\rm{cell}}$ (nm)& 120 & 160 \\
%$a$ (K$/\sqrt{\rm{\mu W}}$) & 8.5 & 7.9 \\
%$L_{\rm{th}}$ (nm) & 264 & 500 \\
%$R_{\rm{in}}T_e$ ($\rm{K^2/(m\cdot W)}$) ~~~~& $2.7\times10^{14}$ & $1.2\times10^{14}$ \\
%$R_{\rm{ex}}T_e$ ($\rm{K^2\cdot m/W}$)& $9.5$ & $15.6$ \\
%$R^{\rm{est}}_{\rm{in}}$ ($\rm{K^2/(m\cdot W)}$) &$4.0\times10^{14}$&$3.5\times10^{14}$\\
%\hline\hline
%\end{tabular}
%\caption{The device-dependent parameters calculated from the experimental results assuming $\beta=1$, and order estimation results (the last row).}
%\label{table}
%\end{table}
%
%Table~\ref{table} presents the device-dependent parameters estimated from the experiments, where we compare devices A and B. 
%To evaluate the thermal resistances, we use the relations of $R_{\rm{in}}T_e^{\beta}=a^{\beta+1}/L_{\rm{th}}$ and $R_{\rm{ex}}T_e^{\beta}=a^{\beta+1}L_{\rm{th}}/(\beta+1)$ derived from the definitions of the $a$ and $L_{\rm{th}}$. Here we use $\beta=1$. We analyze these results to the order estimation values of heat inflow resistance $R^{\rm{est}}_{\rm{in}}$ on the basis of the device structures and the thermophysical properties shown in Supplementary Fig.~\ref{fig8}. Note that the order estimation of $R_{\rm{ex}}$ is difficult due to the complicated heat dissipation path.

Using the model of the QD array structure shown in Fig.~2 in the main text analogous to the series resistance, we estimate $R^{\rm{est}}_{\rm{in}}$ as
\begin{equation}
R^{\rm{est}}_{\rm{in}} = \frac{1}{S L_{\rm{cell}}}\left( \frac{L_{\rm{PolySi}}}{\kappa_{\rm{PolySi}}} + \frac{L_{\rm{SiO_2}}}{\kappa_{\rm{SiO_2}}}\right),\label{Rinest}
\end{equation}
where $S$ is the cross-sectional area of the heat inflow path, $L_{\rm{PolySi}}$ ($L_{\rm{SiO_2}}$) is the total length of the polysilicon gates BG and PG (SiO$_2$), where $L_{\rm{cell}} = L_{\rm{PolySi}}+L_{\rm{SiO_2}}$, and $\kappa_{\rm{PolySi}}$ ($\kappa_{\rm{SiO_2}}$) is the thermal conductivity of the polysilicon gates (SiO$_2$). We use $S=10^{-13}$ $\rm{m}^2$, where $S$ is calculated as the product of the width ($W$) and height ($H$) of the heat inflow path and we use $W=1~\rm{\mu m}$ and $H=0.1~\rm{\mu m}$. From the device structure, we use $L_{\rm{PolySi}}=90$ nm for device A and $L_{\rm{PolySi}}=130$ nm for device B, while $L_{\rm{SiO_2}}=30$ nm for both devices. From the thermophysical properties shown in Fig.~\ref{fig8}, we use $\kappa_{\rm{PolySi}}=0.05~\rm{W/(m\cdot K)}$ and $\kappa_{\rm{SiO_2}}=0.01~\rm{W/(m\cdot K)}$ assuming around 1 K. We obtain the estimation results as $R^{\rm{est}}_{\rm{in}}=4.0\times10^{14}$ for device A and $R^{\rm{est}}_{\rm{in}}=3.5\times10^{14}$ for device B at $T_e=1$ K, as shown in Table~1 in the main text. The lower $R_{\rm{in}}$ for device B than for device A is consistent with the experimental trend. This result demonstrates the validity of our model. Additionally, the order of $R^{\rm{est}}_{\rm{in}}$ values are comparable to the experimental results at $T_e=1$ K, validating the formulation in Eq.~(\ref{Rinest}) at least for the order estimation. We also expect that by improving this simple model or combining it with different approaches, e.g., finite element simulations, we can develop a framework for predicting thermal characteristics more reliably and accurately.

\section{Finite and descrete thermal circuit model}
%############################################
We present a finite and discrete thermal circuit model, analyzed using the ABCD matrix method.  
The ABCD matrix (also known as the transmission matrix) describes the relationship between input and output temperatures (analogous to voltages) and powers (analogous to currents) in a series of two-port networks shown in Supplementary Fig.~\ref{fig2}(a) as
\begin{align}
\begin{bmatrix}
T_e[0] \\
P[0]
\end{bmatrix}
&=
\begin{bmatrix}
1+\frac{R_{\rm{in}}[1]}{R_{\rm{ex}}[1]} & R_{\rm{in}}[1] \\
\frac{1}{R_{\rm{ex}}[1]} & 1
\end{bmatrix}
\cdots
\begin{bmatrix}
1+\frac{R_{\rm{in}}[n]}{R_{\rm{ex}}[n]} & R_{\rm{in}}[n] \\
\frac{1}{R_{\rm{ex}}[n]} & 1
\end{bmatrix}
\begin{bmatrix}
T_e[n] \\
P[n]
\end{bmatrix}
\notag\\
&=
\begin{bmatrix}
1+\frac{a_{\rm{in}}}{a_{\rm{ex}}} & \frac{a_{\rm{in}}}{T_e[1]^{\beta}} \\
\frac{T_e[1]^{\beta}}{a_{\rm{ex}}} & 1
\end{bmatrix}
\cdots
\begin{bmatrix}
1+\frac{a_{\rm{in}}}{a_{\rm{ex}}} & \frac{a_{\rm{in}}}{T_e[n]^{\beta}} \\
\frac{T_e[n]^{\beta}}{a_{\rm{ex}}} & 1
\end{bmatrix}
\begin{bmatrix}
T_e[n] \\
P[n]
\end{bmatrix}
\end{align}
where we assume $T_e[i] \gg T_{\rm{base}}$ and $\beta_{\rm{in}} \approx \beta_{\rm{ex}} \, (\equiv \beta)$ similar to the assumptions in the main text.
Here, we apply the boundary conditions: $P[0] = P_0$ and $P[n] = 0$.  
Then, we numerically solve for $T_e[i]$ where $i = 0, \ldots, n$, given $P_0$.  
Supplementary Fig.~\ref{fig2}(b-d) illustrate the case for $n = 4$ and $\beta=2$.  
The results are well fitted by Eq.~(5) in the main text.
In this figure, we show the results for $\beta=2$, but we have confirmed that it works well for any value of $\beta\geq 0$.  
These results show that even in a finite and discrete circuit, the parameter dependence can be modeled by Eq.~(5) in the main text.
\begin{figure}[h]
\centering
\includegraphics[width=80mm]{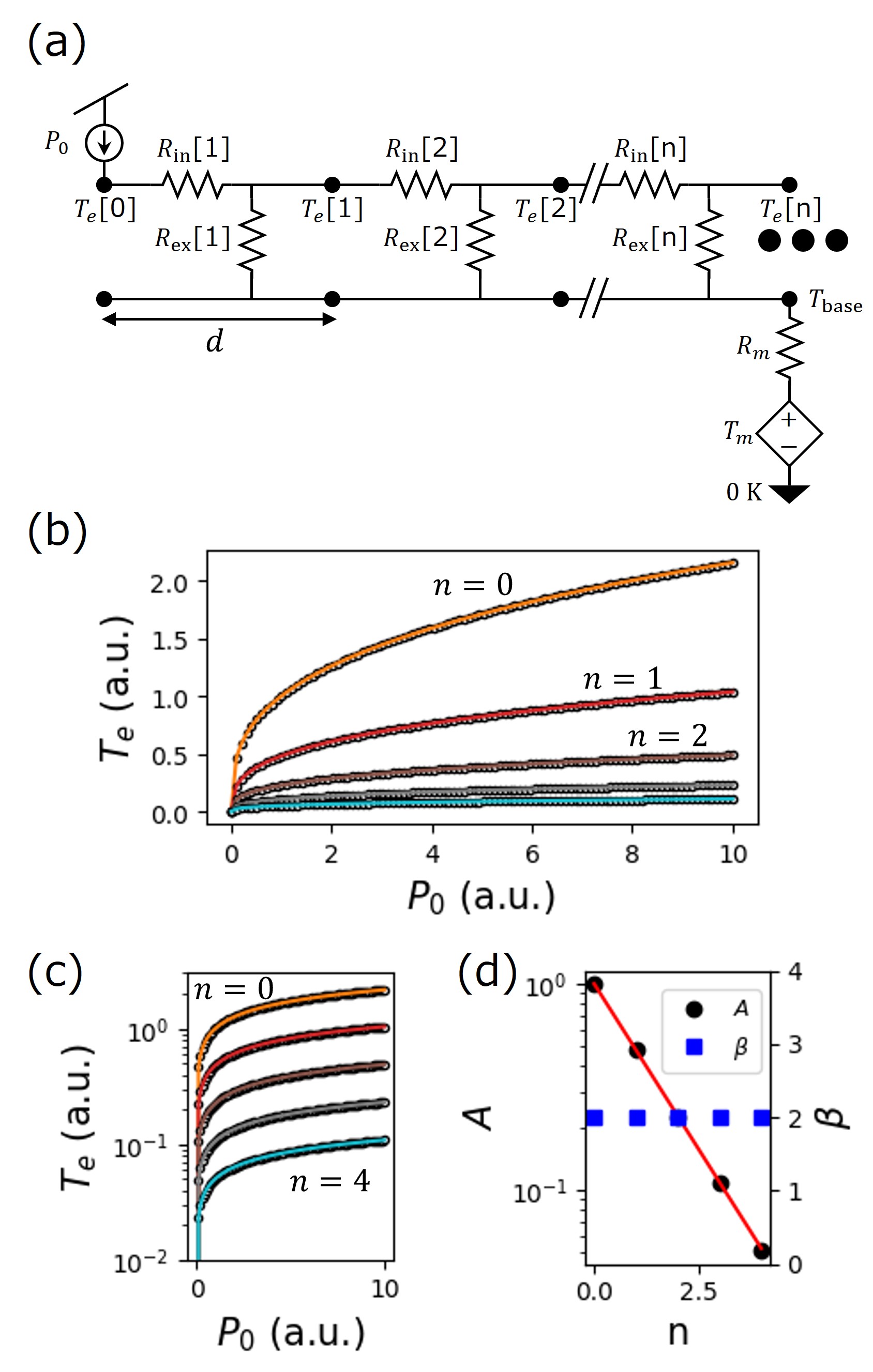}
\caption{
(a) Finite and discrete thermal circuit model, and (b–d) calculation results for the case of $n = 4$, where $\beta$ is set to 2, and $a_{\rm{in}} = a_{\rm{ex}} = 1$.  
(b) Linear plots showing the dependence of $T_e$ on $P_0$, and (c) the corresponding log plots.  
The colored lines represent the fitting curves based on Eq.~(5) in the main text.  
(d) Best-fit results for $A = ae^{-x/L_{\rm{th}}}$ and $\beta$, where $x = nd$, implying $A = ae^{d/L_{\rm{th}}}e^{-n} \propto e^{-n}$.}
\label{fig2}
\end{figure}

%\section{Calibration for electron temperature extraction}
%We measure the extracted electron temperature $T_e$ varying MXC temperature.
%$T_e$ is extracted by using Eq.~(8) in the main taxt, where lever arm $\alpha$ is extracted by Coulomb blockade measurement.

\section{Validity of CBT setup}
\label{app1}
\begin{figure}
\includegraphics[width=50mm]{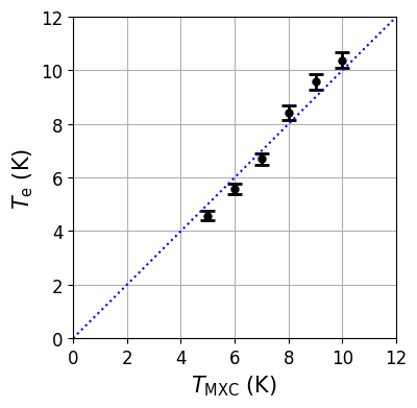}%
\caption{\label{fig5}Estimated electron temperature $T_e$ as a function of the temperature of the mixing chamber $T_{\rm{MXC}}$. Error bar shows the standard deviation error of $T_e$ estimate.}%, where the lever arm was estimated as $\alpha = 0.04$ by obtaining the Coulomb diamond.
\end{figure}

We evaluate the validity of the CBT setup. In this measurement, we use the device A with $L_{\rm{cell}}=120$ nm. Supplementary Fig.~\ref{fig5} shows the estimated electron temperature $T_e$ obtained by CBT under various temperatures of the mixing chamber $T_{\rm{MXC}}$ in the dilution refrigerator.
In the high-temperature regime above 5 K, $T_{\rm{MXC}}$ and $T_e$ are expected to be the same. As expected, the results of $T_{\rm{MXC}}=T_e$ were obtained, confirming that the temperature estimation using this CBT measurement setup is valid.
For the higher temperature regime of 10–-30 K shown in Fig. 5(b) and 5(c) in the main text, the validity is supported by extrapolation from the confirmed data.

\begin{figure}
\includegraphics[width=85mm]{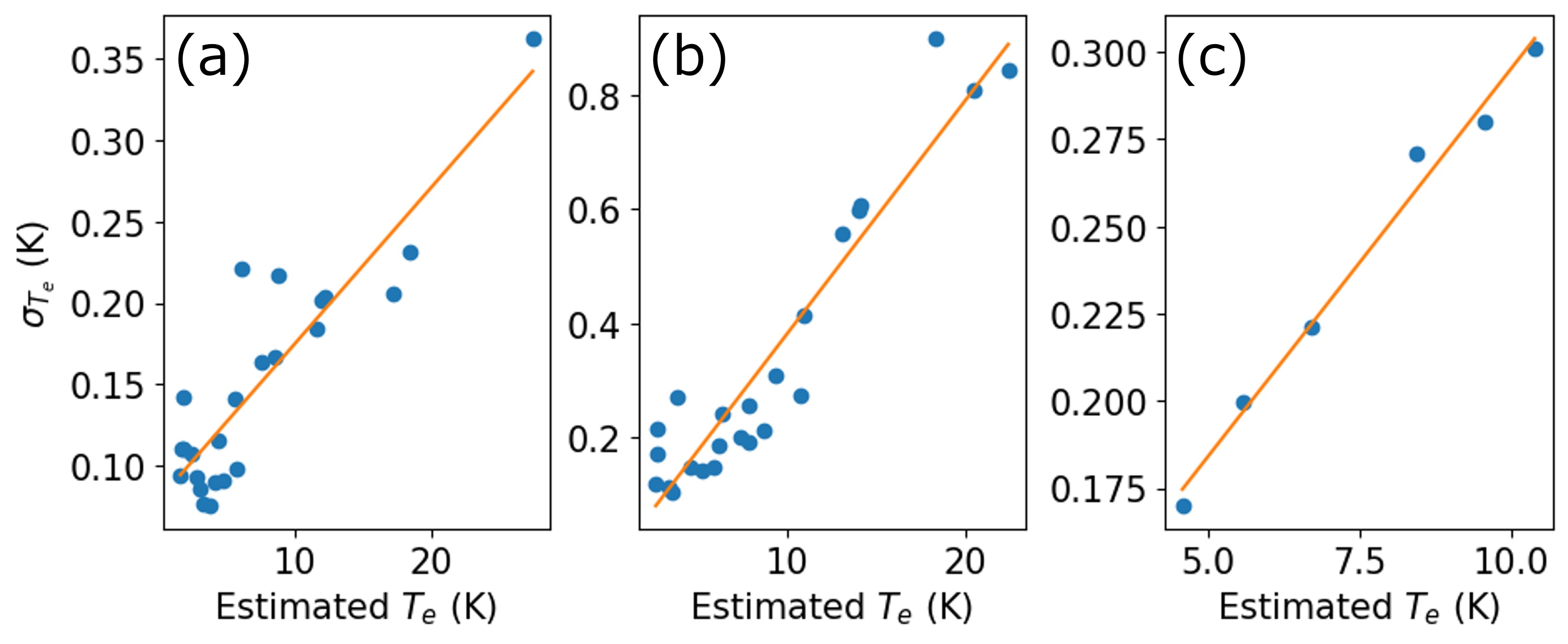}%
\caption{\label{fig6}Standard deviation error of $T_e$ estimate $\sigma_{T_e}$ as a function of estimated $T_e$ according to (a) Fig. 5(b) and (b) Fig. 5(c) in the main text, and (c) Supplementary Fig.~\ref{fig5}, respectevely.}%, where the lever arm was estimated as $\alpha = 0.04$ by obtaining the Coulomb diamond.
\end{figure}

We discuss the uncertainty in extracting $T_e$ using CBT.  
Supplementary Fig.~\ref{fig6} illustrates the relationship between the estimated $T_e$ and the associated error for three cases, corresponding to Fig.~5(a) and Fig.~5(b) in the main text, and Supplementary Fig.~\ref{fig5}, respectively.  
The error appears to scale proportionally with the estimated $T_e$, suggesting that the relative error rate, $\sigma_{T_e} / T_e$, remains approximately constant throughout this measurement.
%The proportionality of the error is also explained by following equation:
%\begin{equation}
%\sqrt{\int_{-\infty}^{\infty} \left( \frac{d}{dT_e} \left[ \cosh^{-2}\left(\frac{\alpha eV}{2.5 k_B T_e}\right) \right] \right)^2 dV} \propto \frac{1}{\sqrt{T_e}},
%\end{equation}
%where the left-hand side is the error index in the fitting function shown in Eq.~8 when $T_e$ is slightly displaced.

\section*{References}
\bibliography{utsugi-paper}

\begin{thebibliography}{45}%
\makeatletter
\providecommand \@ifxundefined [1]{%
 \@ifx{#1\undefined}
}%
\providecommand \@ifnum [1]{%
 \ifnum #1\expandafter \@firstoftwo
 \else \expandafter \@secondoftwo
 \fi
}%
\providecommand \@ifx [1]{%
 \ifx #1\expandafter \@firstoftwo
 \else \expandafter \@secondoftwo
 \fi
}%
\providecommand \natexlab [1]{#1}%
\providecommand \enquote  [1]{``#1''}%
\providecommand \bibnamefont  [1]{#1}%
\providecommand \bibfnamefont [1]{#1}%
\providecommand \citenamefont [1]{#1}%
\providecommand \href@noop [0]{\@secondoftwo}%
\providecommand \href [0]{\begingroup \@sanitize@url \@href}%
\providecommand \@href[1]{\@@startlink{#1}\@@href}%
\providecommand \@@href[1]{\endgroup#1\@@endlink}%
\providecommand \@sanitize@url [0]{\catcode `\\12\catcode `\$12\catcode
  `\&12\catcode `\#12\catcode `\^12\catcode `\_12\catcode `\%12\relax}%
\providecommand \@@startlink[1]{}%
\providecommand \@@endlink[0]{}%
\providecommand \url  [0]{\begingroup\@sanitize@url \@url }%
\providecommand \@url [1]{\endgroup\@href {#1}{\urlprefix }}%
\providecommand \urlprefix  [0]{URL }%
\providecommand \Eprint [0]{\href }%
\providecommand \doibase [0]{http://dx.doi.org/}%
\providecommand \selectlanguage [0]{\@gobble}%
\providecommand \bibinfo  [0]{\@secondoftwo}%
\providecommand \bibfield  [0]{\@secondoftwo}%
\providecommand \translation [1]{[#1]}%
\providecommand \BibitemOpen [0]{}%
\providecommand \bibitemStop [0]{}%
\providecommand \bibitemNoStop [0]{.\EOS\space}%
\providecommand \EOS [0]{\spacefactor3000\relax}%
\providecommand \BibitemShut  [1]{\csname bibitem#1\endcsname}%
\let\auto@bib@innerbib\@empty
%</preamble>
\bibitem [{\citenamefont {Burkard}\ \emph {et~al.}(2023)\citenamefont
  {Burkard}, \citenamefont {Ladd}, \citenamefont {Pan}, \citenamefont
  {Nichol},\ and\ \citenamefont {Petta}}]{Burkard2023}%
  \BibitemOpen
  \bibfield  {author} {\bibinfo {author} {\bibfnamefont {G.}~\bibnamefont
  {Burkard}}, \bibinfo {author} {\bibfnamefont {T.~D.}\ \bibnamefont {Ladd}},
  \bibinfo {author} {\bibfnamefont {A.}~\bibnamefont {Pan}}, \bibinfo {author}
  {\bibfnamefont {J.~M.}\ \bibnamefont {Nichol}}, \ and\ \bibinfo {author}
  {\bibfnamefont {J.~R.}\ \bibnamefont {Petta}},\ }\bibfield  {title} {\enquote
  {\bibinfo {title} {Semiconductor spin qubits},}\ }\href@noop {} {\bibfield
  {journal} {\bibinfo  {journal} {Reviews of Modern Physics}\ }\textbf
  {\bibinfo {volume} {95}},\ \bibinfo {pages} {025003} (\bibinfo {year}
  {2023})}\BibitemShut {NoStop}%
\bibitem [{\citenamefont {Kodera}(2024)}]{kodera2024trends}%
  \BibitemOpen
  \bibfield  {author} {\bibinfo {author} {\bibfnamefont {T.}~\bibnamefont
  {Kodera}},\ }\bibfield  {title} {\enquote {\bibinfo {title} {Trends and
  prospects for semiconductor qubit research},}\ }\href@noop {} {\bibfield
  {journal} {\bibinfo  {journal} {JSAP Review}\ }\textbf {\bibinfo {volume}
  {2024}},\ \bibinfo {pages} {240101} (\bibinfo {year} {2024})}\BibitemShut
  {NoStop}%
\bibitem [{\citenamefont {Vandersypen}\ \emph {et~al.}(2017)\citenamefont
  {Vandersypen}, \citenamefont {Bluhm}, \citenamefont {Clarke}, \citenamefont
  {Dzurak}, \citenamefont {Ishihara}, \citenamefont {Morello}, \citenamefont
  {Reilly}, \citenamefont {Schreiber},\ and\ \citenamefont
  {Veldhorst}}]{Vandersypen2017}%
  \BibitemOpen
  \bibfield  {author} {\bibinfo {author} {\bibfnamefont {L.~M.~K.}\
  \bibnamefont {Vandersypen}}, \bibinfo {author} {\bibfnamefont
  {H.}~\bibnamefont {Bluhm}}, \bibinfo {author} {\bibfnamefont {J.~S.}\
  \bibnamefont {Clarke}}, \bibinfo {author} {\bibfnamefont {A.~S.}\
  \bibnamefont {Dzurak}}, \bibinfo {author} {\bibfnamefont {R.}~\bibnamefont
  {Ishihara}}, \bibinfo {author} {\bibfnamefont {A.}~\bibnamefont {Morello}},
  \bibinfo {author} {\bibfnamefont {D.~J.}\ \bibnamefont {Reilly}}, \bibinfo
  {author} {\bibfnamefont {L.~R.}\ \bibnamefont {Schreiber}}, \ and\ \bibinfo
  {author} {\bibfnamefont {M.}~\bibnamefont {Veldhorst}},\ }\bibfield  {title}
  {\enquote {\bibinfo {title} {Interfacing spin qubits in quantum dots and
  donors -hot, dense, and coherent},}\ }\href@noop {} {\bibfield  {journal}
  {\bibinfo  {journal} {NPJ Quantum Inf.}\ }\textbf {\bibinfo {volume} {3}},\
  \bibinfo {pages} {34} (\bibinfo {year} {2017})}\BibitemShut {NoStop}%
\bibitem [{\citenamefont {Veldhorst}\ \emph {et~al.}(2017)\citenamefont
  {Veldhorst}, \citenamefont {Eenink}, \citenamefont {Yang},\ and\
  \citenamefont {Dzurak}}]{Veldhorst2017}%
  \BibitemOpen
  \bibfield  {author} {\bibinfo {author} {\bibfnamefont {M.}~\bibnamefont
  {Veldhorst}}, \bibinfo {author} {\bibfnamefont {H.~G.~J.}\ \bibnamefont
  {Eenink}}, \bibinfo {author} {\bibfnamefont {C.-H.}\ \bibnamefont {Yang}}, \
  and\ \bibinfo {author} {\bibfnamefont {A.~S.}\ \bibnamefont {Dzurak}},\
  }\bibfield  {title} {\enquote {\bibinfo {title} {Silicon cmos architecture
  for a spin-based quantum computer},}\ }\href@noop {} {\bibfield  {journal}
  {\bibinfo  {journal} {Nature communications}\ }\textbf {\bibinfo {volume}
  {8}},\ \bibinfo {pages} {1766} (\bibinfo {year} {2017})}\BibitemShut
  {NoStop}%
\bibitem [{\citenamefont {Borsoi}\ \emph {et~al.}(2024)\citenamefont {Borsoi},
  \citenamefont {Hendrickx}, \citenamefont {John}, \citenamefont {Meyer},
  \citenamefont {Motz}, \citenamefont {van Riggelen}, \citenamefont {Sammak},
  \citenamefont {de~Snoo}, \citenamefont {Scappucci},\ and\ \citenamefont
  {Veldhorst}}]{Borsoi2022}%
  \BibitemOpen
  \bibfield  {author} {\bibinfo {author} {\bibfnamefont {F.}~\bibnamefont
  {Borsoi}}, \bibinfo {author} {\bibfnamefont {N.~W.}\ \bibnamefont
  {Hendrickx}}, \bibinfo {author} {\bibfnamefont {V.}~\bibnamefont {John}},
  \bibinfo {author} {\bibfnamefont {M.}~\bibnamefont {Meyer}}, \bibinfo
  {author} {\bibfnamefont {S.}~\bibnamefont {Motz}}, \bibinfo {author}
  {\bibfnamefont {F.}~\bibnamefont {van Riggelen}}, \bibinfo {author}
  {\bibfnamefont {A.}~\bibnamefont {Sammak}}, \bibinfo {author} {\bibfnamefont
  {S.~L.}\ \bibnamefont {de~Snoo}}, \bibinfo {author} {\bibfnamefont
  {G.}~\bibnamefont {Scappucci}}, \ and\ \bibinfo {author} {\bibfnamefont
  {M.}~\bibnamefont {Veldhorst}},\ }\bibfield  {title} {\enquote {\bibinfo
  {title} {Shared control of a 16 semiconductor quantum dot crossbar array},}\
  }\href@noop {} {\bibfield  {journal} {\bibinfo  {journal} {Nature
  Nanotechnology}\ }\textbf {\bibinfo {volume} {19}},\ \bibinfo {pages}
  {21--27} (\bibinfo {year} {2024})}\BibitemShut {NoStop}%
\bibitem [{\citenamefont {Li}\ \emph {et~al.}(2018)\citenamefont {Li},
  \citenamefont {Petit}, \citenamefont {Franke}, \citenamefont {Dehollain},
  \citenamefont {Helsen}, \citenamefont {Steudtner}, \citenamefont {Thomas},
  \citenamefont {Yoscovits}, \citenamefont {Singh}, \citenamefont {Wehner}
  \emph {et~al.}}]{Li2018}%
  \BibitemOpen
  \bibfield  {author} {\bibinfo {author} {\bibfnamefont {R.}~\bibnamefont
  {Li}}, \bibinfo {author} {\bibfnamefont {L.}~\bibnamefont {Petit}}, \bibinfo
  {author} {\bibfnamefont {D.~P.}\ \bibnamefont {Franke}}, \bibinfo {author}
  {\bibfnamefont {J.~P.}\ \bibnamefont {Dehollain}}, \bibinfo {author}
  {\bibfnamefont {J.}~\bibnamefont {Helsen}}, \bibinfo {author} {\bibfnamefont
  {M.}~\bibnamefont {Steudtner}}, \bibinfo {author} {\bibfnamefont {N.~K.}\
  \bibnamefont {Thomas}}, \bibinfo {author} {\bibfnamefont {Z.~R.}\
  \bibnamefont {Yoscovits}}, \bibinfo {author} {\bibfnamefont {K.~J.}\
  \bibnamefont {Singh}}, \bibinfo {author} {\bibfnamefont {S.}~\bibnamefont
  {Wehner}},  \emph {et~al.},\ }\bibfield  {title} {\enquote {\bibinfo {title}
  {A crossbar network for silicon quantum dot qubits},}\ }\href@noop {}
  {\bibfield  {journal} {\bibinfo  {journal} {Science advances}\ }\textbf
  {\bibinfo {volume} {4}},\ \bibinfo {pages} {eaar3960} (\bibinfo {year}
  {2018})}\BibitemShut {NoStop}%
\bibitem [{\citenamefont {Philips}\ \emph {et~al.}(2022)\citenamefont
  {Philips}, \citenamefont {Madzik}, \citenamefont {Amitonov}, \citenamefont
  {Snoo}, \citenamefont {Russ}, \citenamefont {Kalhor}, \citenamefont {Volk},
  \citenamefont {Lawrie}, \citenamefont {Brousse}, \citenamefont {Tryputen}
  \emph {et~al.}}]{Philips2022}%
  \BibitemOpen
  \bibfield  {author} {\bibinfo {author} {\bibfnamefont {S.~G.~J.}\
  \bibnamefont {Philips}}, \bibinfo {author} {\bibfnamefont {M.~T.}\
  \bibnamefont {Madzik}}, \bibinfo {author} {\bibfnamefont {S.~V.}\
  \bibnamefont {Amitonov}}, \bibinfo {author} {\bibfnamefont {S.~L.~D.}\
  \bibnamefont {Snoo}}, \bibinfo {author} {\bibfnamefont {M.}~\bibnamefont
  {Russ}}, \bibinfo {author} {\bibfnamefont {N.}~\bibnamefont {Kalhor}},
  \bibinfo {author} {\bibfnamefont {C.}~\bibnamefont {Volk}}, \bibinfo {author}
  {\bibfnamefont {W.~I.~L.}\ \bibnamefont {Lawrie}}, \bibinfo {author}
  {\bibfnamefont {D.}~\bibnamefont {Brousse}}, \bibinfo {author} {\bibfnamefont
  {L.}~\bibnamefont {Tryputen}},  \emph {et~al.},\ }\bibfield  {title}
  {\enquote {\bibinfo {title} {Universal control of a six-qubit quantum
  processor in silicon},}\ }\href@noop {} {\bibfield  {journal} {\bibinfo
  {journal} {Nature}\ }\textbf {\bibinfo {volume} {609}},\ \bibinfo {pages}
  {919--924} (\bibinfo {year} {2022})}\BibitemShut {NoStop}%
\bibitem [{\citenamefont {Lee}\ \emph {et~al.}(2020)\citenamefont {Lee},
  \citenamefont {Tsuchiya}, \citenamefont {Shinkai}, \citenamefont {Kanno},
  \citenamefont {Mine}, \citenamefont {Takahama}, \citenamefont {Mizokuchi},
  \citenamefont {Kodera}, \citenamefont {Hisamoto},\ and\ \citenamefont
  {Mizuno}}]{Lee2020}%
  \BibitemOpen
  \bibfield  {author} {\bibinfo {author} {\bibfnamefont {N.}~\bibnamefont
  {Lee}}, \bibinfo {author} {\bibfnamefont {R.}~\bibnamefont {Tsuchiya}},
  \bibinfo {author} {\bibfnamefont {G.}~\bibnamefont {Shinkai}}, \bibinfo
  {author} {\bibfnamefont {Y.}~\bibnamefont {Kanno}}, \bibinfo {author}
  {\bibfnamefont {T.}~\bibnamefont {Mine}}, \bibinfo {author} {\bibfnamefont
  {T.}~\bibnamefont {Takahama}}, \bibinfo {author} {\bibfnamefont
  {R.}~\bibnamefont {Mizokuchi}}, \bibinfo {author} {\bibfnamefont
  {T.}~\bibnamefont {Kodera}}, \bibinfo {author} {\bibfnamefont
  {D.}~\bibnamefont {Hisamoto}}, \ and\ \bibinfo {author} {\bibfnamefont
  {H.}~\bibnamefont {Mizuno}},\ }\bibfield  {title} {\enquote {\bibinfo {title}
  {Enhancing electrostatic coupling in silicon quantum dot array by dual gate
  oxide thickness for large-scale integration},}\ }\href@noop {} {\bibfield
  {journal} {\bibinfo  {journal} {Appl. Phys. Lett.}\ }\textbf {\bibinfo
  {volume} {116}},\ \bibinfo {pages} {162106} (\bibinfo {year}
  {2020})}\BibitemShut {NoStop}%
\bibitem [{\citenamefont {Lee}\ \emph {et~al.}(2022)\citenamefont {Lee},
  \citenamefont {Tsuchiya}, \citenamefont {Kanno}, \citenamefont {Mine},
  \citenamefont {Sasago}, \citenamefont {Shinkai}, \citenamefont {Mizokuchi},
  \citenamefont {Yoneda}, \citenamefont {Kodera}, \citenamefont {Yoshimura},
  \citenamefont {Saito}, \citenamefont {Hisamoto},\ and\ \citenamefont
  {Mizuno}}]{Lee2022}%
  \BibitemOpen
  \bibfield  {author} {\bibinfo {author} {\bibfnamefont {N.}~\bibnamefont
  {Lee}}, \bibinfo {author} {\bibfnamefont {R.}~\bibnamefont {Tsuchiya}},
  \bibinfo {author} {\bibfnamefont {Y.}~\bibnamefont {Kanno}}, \bibinfo
  {author} {\bibfnamefont {T.}~\bibnamefont {Mine}}, \bibinfo {author}
  {\bibfnamefont {Y.}~\bibnamefont {Sasago}}, \bibinfo {author} {\bibfnamefont
  {G.}~\bibnamefont {Shinkai}}, \bibinfo {author} {\bibfnamefont
  {R.}~\bibnamefont {Mizokuchi}}, \bibinfo {author} {\bibfnamefont
  {J.}~\bibnamefont {Yoneda}}, \bibinfo {author} {\bibfnamefont
  {T.}~\bibnamefont {Kodera}}, \bibinfo {author} {\bibfnamefont
  {C.}~\bibnamefont {Yoshimura}}, \bibinfo {author} {\bibfnamefont
  {S.}~\bibnamefont {Saito}}, \bibinfo {author} {\bibfnamefont
  {D.}~\bibnamefont {Hisamoto}}, \ and\ \bibinfo {author} {\bibfnamefont
  {H.}~\bibnamefont {Mizuno}},\ }\bibfield  {title} {\enquote {\bibinfo {title}
  {16 x 8 quantum dot array operation at cryogenic temperatures},}\ }\href
  {\doibase 10.35848/1347-4065/ac4c07} {\bibfield  {journal} {\bibinfo
  {journal} {Jpn. J. Appl. Phys.}\ }\textbf {\bibinfo {volume} {61}},\ \bibinfo
  {pages} {1040} (\bibinfo {year} {2022})}\BibitemShut {NoStop}%
\bibitem [{\citenamefont {Huang}\ \emph {et~al.}(2024)\citenamefont {Huang},
  \citenamefont {Su}, \citenamefont {Lim}, \citenamefont {Feng}, \citenamefont
  {van Straaten}, \citenamefont {Severin}, \citenamefont {Gilbert},
  \citenamefont {Dumoulin~Stuyck}, \citenamefont {Tanttu}, \citenamefont
  {Serrano} \emph {et~al.}}]{huang2024high}%
  \BibitemOpen
  \bibfield  {author} {\bibinfo {author} {\bibfnamefont {J.~Y.}\ \bibnamefont
  {Huang}}, \bibinfo {author} {\bibfnamefont {R.~Y.}\ \bibnamefont {Su}},
  \bibinfo {author} {\bibfnamefont {W.~H.}\ \bibnamefont {Lim}}, \bibinfo
  {author} {\bibfnamefont {M.}~\bibnamefont {Feng}}, \bibinfo {author}
  {\bibfnamefont {B.}~\bibnamefont {van Straaten}}, \bibinfo {author}
  {\bibfnamefont {B.}~\bibnamefont {Severin}}, \bibinfo {author} {\bibfnamefont
  {W.}~\bibnamefont {Gilbert}}, \bibinfo {author} {\bibfnamefont
  {N.}~\bibnamefont {Dumoulin~Stuyck}}, \bibinfo {author} {\bibfnamefont
  {T.}~\bibnamefont {Tanttu}}, \bibinfo {author} {\bibfnamefont
  {S.}~\bibnamefont {Serrano}},  \emph {et~al.},\ }\bibfield  {title} {\enquote
  {\bibinfo {title} {High-fidelity spin qubit operation and algorithmic
  initialization above 1 k},}\ }\href@noop {} {\bibfield  {journal} {\bibinfo
  {journal} {Nature}\ }\textbf {\bibinfo {volume} {627}},\ \bibinfo {pages}
  {772--777} (\bibinfo {year} {2024})}\BibitemShut {NoStop}%
\bibitem [{\citenamefont {Krinner}\ \emph {et~al.}(2019)\citenamefont
  {Krinner}, \citenamefont {Storz}, \citenamefont {Kurpiers}, \citenamefont
  {Magnard}, \citenamefont {Heinsoo}, \citenamefont {Keller}, \citenamefont
  {Lütolf}, \citenamefont {Eichler},\ and\ \citenamefont
  {Wallraff}}]{Krinner2019}%
  \BibitemOpen
  \bibfield  {author} {\bibinfo {author} {\bibfnamefont {S.}~\bibnamefont
  {Krinner}}, \bibinfo {author} {\bibfnamefont {S.}~\bibnamefont {Storz}},
  \bibinfo {author} {\bibfnamefont {P.}~\bibnamefont {Kurpiers}}, \bibinfo
  {author} {\bibfnamefont {P.}~\bibnamefont {Magnard}}, \bibinfo {author}
  {\bibfnamefont {J.}~\bibnamefont {Heinsoo}}, \bibinfo {author} {\bibfnamefont
  {R.}~\bibnamefont {Keller}}, \bibinfo {author} {\bibfnamefont
  {J.}~\bibnamefont {Lütolf}}, \bibinfo {author} {\bibfnamefont
  {C.}~\bibnamefont {Eichler}}, \ and\ \bibinfo {author} {\bibfnamefont
  {A.}~\bibnamefont {Wallraff}},\ }\bibfield  {title} {\enquote {\bibinfo
  {title} {Engineering cryogenic setups for 100-qubit scale superconducting
  circuit systems},}\ }\href {\doibase 10.1140/epjqt/s40507-019-0072-0}
  {\bibfield  {journal} {\bibinfo  {journal} {EPJ Quantum Technol.}\ }\textbf
  {\bibinfo {volume} {6}},\ \bibinfo {pages} {2} (\bibinfo {year}
  {2019})}\BibitemShut {NoStop}%
\bibitem [{\citenamefont {Savin}\ \emph {et~al.}(2006)\citenamefont {Savin},
  \citenamefont {Pekola}, \citenamefont {Averin},\ and\ \citenamefont
  {Semenov}}]{Savin2006}%
  \BibitemOpen
  \bibfield  {author} {\bibinfo {author} {\bibfnamefont {A.~M.}\ \bibnamefont
  {Savin}}, \bibinfo {author} {\bibfnamefont {J.~P.}\ \bibnamefont {Pekola}},
  \bibinfo {author} {\bibfnamefont {D.~V.}\ \bibnamefont {Averin}}, \ and\
  \bibinfo {author} {\bibfnamefont {V.~K.}\ \bibnamefont {Semenov}},\
  }\bibfield  {title} {\enquote {\bibinfo {title} {Thermal budget of
  superconducting digital circuits at subkelvin temperatures},}\ }\href@noop {}
  {\bibfield  {journal} {\bibinfo  {journal} {Journal of applied physics}\
  }\textbf {\bibinfo {volume} {99}} (\bibinfo {year} {2006})}\BibitemShut
  {NoStop}%
\bibitem [{\citenamefont {Kawakami}\ \emph {et~al.}(2013)\citenamefont
  {Kawakami}, \citenamefont {Scarlino}, \citenamefont {Schreiber},
  \citenamefont {Prance}, \citenamefont {Savage}, \citenamefont {Lagally},
  \citenamefont {Eriksson},\ and\ \citenamefont {Vandersypen}}]{Kawakami2013}%
  \BibitemOpen
  \bibfield  {author} {\bibinfo {author} {\bibfnamefont {E.}~\bibnamefont
  {Kawakami}}, \bibinfo {author} {\bibfnamefont {P.}~\bibnamefont {Scarlino}},
  \bibinfo {author} {\bibfnamefont {L.~R.}\ \bibnamefont {Schreiber}}, \bibinfo
  {author} {\bibfnamefont {J.~R.}\ \bibnamefont {Prance}}, \bibinfo {author}
  {\bibfnamefont {D.~E.}\ \bibnamefont {Savage}}, \bibinfo {author}
  {\bibfnamefont {M.~G.}\ \bibnamefont {Lagally}}, \bibinfo {author}
  {\bibfnamefont {M.~A.}\ \bibnamefont {Eriksson}}, \ and\ \bibinfo {author}
  {\bibfnamefont {L.~M.~K.}\ \bibnamefont {Vandersypen}},\ }\bibfield  {title}
  {\enquote {\bibinfo {title} {Excitation of a si/sige quantum dot using an
  on-chip microwave antenna},}\ }\href@noop {} {\bibfield  {journal} {\bibinfo
  {journal} {Applied Physics Letters}\ }\textbf {\bibinfo {volume} {103}}
  (\bibinfo {year} {2013})}\BibitemShut {NoStop}%
\bibitem [{\citenamefont {Takeda}\ \emph {et~al.}(2018)\citenamefont {Takeda},
  \citenamefont {Yoneda}, \citenamefont {Otsuka}, \citenamefont {Nakajima},
  \citenamefont {Delbecq}, \citenamefont {Allison}, \citenamefont {Hoshi},
  \citenamefont {Usami}, \citenamefont {Itoh}, \citenamefont {Oda} \emph
  {et~al.}}]{takeda2018optimized}%
  \BibitemOpen
  \bibfield  {author} {\bibinfo {author} {\bibfnamefont {K.}~\bibnamefont
  {Takeda}}, \bibinfo {author} {\bibfnamefont {J.}~\bibnamefont {Yoneda}},
  \bibinfo {author} {\bibfnamefont {T.}~\bibnamefont {Otsuka}}, \bibinfo
  {author} {\bibfnamefont {T.}~\bibnamefont {Nakajima}}, \bibinfo {author}
  {\bibfnamefont {M.}~\bibnamefont {Delbecq}}, \bibinfo {author} {\bibfnamefont
  {G.}~\bibnamefont {Allison}}, \bibinfo {author} {\bibfnamefont
  {Y.}~\bibnamefont {Hoshi}}, \bibinfo {author} {\bibfnamefont
  {N.}~\bibnamefont {Usami}}, \bibinfo {author} {\bibfnamefont
  {K.}~\bibnamefont {Itoh}}, \bibinfo {author} {\bibfnamefont {S.}~\bibnamefont
  {Oda}},  \emph {et~al.},\ }\bibfield  {title} {\enquote {\bibinfo {title}
  {Optimized electrical control of a si/sige spin qubit in the presence of an
  induced frequency shift},}\ }\href@noop {} {\bibfield  {journal} {\bibinfo
  {journal} {npj Quantum Information}\ }\textbf {\bibinfo {volume} {4}},\
  \bibinfo {pages} {54} (\bibinfo {year} {2018})}\BibitemShut {NoStop}%
\bibitem [{\citenamefont {Noah}\ \emph {et~al.}(2024)\citenamefont {Noah},
  \citenamefont {Swift}, \citenamefont {De~Kruijf}, \citenamefont {Gomez-Saiz},
  \citenamefont {Morton},\ and\ \citenamefont {Gonzalez-Zalba}}]{Noah2023}%
  \BibitemOpen
  \bibfield  {author} {\bibinfo {author} {\bibfnamefont {G.~M.}\ \bibnamefont
  {Noah}}, \bibinfo {author} {\bibfnamefont {T.~H.}\ \bibnamefont {Swift}},
  \bibinfo {author} {\bibfnamefont {M.}~\bibnamefont {De~Kruijf}}, \bibinfo
  {author} {\bibfnamefont {A.}~\bibnamefont {Gomez-Saiz}}, \bibinfo {author}
  {\bibfnamefont {J.~J.}\ \bibnamefont {Morton}}, \ and\ \bibinfo {author}
  {\bibfnamefont {M.~F.}\ \bibnamefont {Gonzalez-Zalba}},\ }\bibfield  {title}
  {\enquote {\bibinfo {title} {Cmos on-chip thermometry at deep cryogenic
  temperatures},}\ }\href@noop {} {\bibfield  {journal} {\bibinfo  {journal}
  {Applied Physics Reviews}\ }\textbf {\bibinfo {volume} {11}} (\bibinfo {year}
  {2024})}\BibitemShut {NoStop}%
\bibitem [{\citenamefont {Undseth}\ \emph {et~al.}(2023)\citenamefont
  {Undseth}, \citenamefont {Pietx-Casas}, \citenamefont {Raymenants},
  \citenamefont {Mehmandoost}, \citenamefont {Mkdzik}, \citenamefont {Philips},
  \citenamefont {De~Snoo}, \citenamefont {Michalak}, \citenamefont {Amitonov},
  \citenamefont {Tryputen} \emph {et~al.}}]{undseth2023hotter}%
  \BibitemOpen
  \bibfield  {author} {\bibinfo {author} {\bibfnamefont {B.}~\bibnamefont
  {Undseth}}, \bibinfo {author} {\bibfnamefont {O.}~\bibnamefont
  {Pietx-Casas}}, \bibinfo {author} {\bibfnamefont {E.}~\bibnamefont
  {Raymenants}}, \bibinfo {author} {\bibfnamefont {M.}~\bibnamefont
  {Mehmandoost}}, \bibinfo {author} {\bibfnamefont {M.~T.}\ \bibnamefont
  {Mkdzik}}, \bibinfo {author} {\bibfnamefont {S.~G.}\ \bibnamefont {Philips}},
  \bibinfo {author} {\bibfnamefont {S.~L.}\ \bibnamefont {De~Snoo}}, \bibinfo
  {author} {\bibfnamefont {D.~J.}\ \bibnamefont {Michalak}}, \bibinfo {author}
  {\bibfnamefont {S.~V.}\ \bibnamefont {Amitonov}}, \bibinfo {author}
  {\bibfnamefont {L.}~\bibnamefont {Tryputen}},  \emph {et~al.},\ }\bibfield
  {title} {\enquote {\bibinfo {title} {Hotter is easier unexpected temperature
  dependence of spin qubit frequencies},}\ }\href@noop {} {\bibfield  {journal}
  {\bibinfo  {journal} {Physical Review X}\ }\textbf {\bibinfo {volume} {13}},\
  \bibinfo {pages} {041015} (\bibinfo {year} {2023})}\BibitemShut {NoStop}%
\bibitem [{\citenamefont {Xue}\ \emph {et~al.}(2021)\citenamefont {Xue},
  \citenamefont {Patra}, \citenamefont {Dijk}, \citenamefont {Samkharadze},
  \citenamefont {Subramanian}, \citenamefont {Corna}, \citenamefont {Wuetz},
  \citenamefont {Jeon}, \citenamefont {Sheikh}, \citenamefont
  {Juarez-Hernandez} \emph {et~al.}}]{Xue2021}%
  \BibitemOpen
  \bibfield  {author} {\bibinfo {author} {\bibfnamefont {X.}~\bibnamefont
  {Xue}}, \bibinfo {author} {\bibfnamefont {B.}~\bibnamefont {Patra}}, \bibinfo
  {author} {\bibfnamefont {J.~P. G.~V.}\ \bibnamefont {Dijk}}, \bibinfo
  {author} {\bibfnamefont {N.}~\bibnamefont {Samkharadze}}, \bibinfo {author}
  {\bibfnamefont {S.}~\bibnamefont {Subramanian}}, \bibinfo {author}
  {\bibfnamefont {A.}~\bibnamefont {Corna}}, \bibinfo {author} {\bibfnamefont
  {B.~P.}\ \bibnamefont {Wuetz}}, \bibinfo {author} {\bibfnamefont
  {C.}~\bibnamefont {Jeon}}, \bibinfo {author} {\bibfnamefont {F.}~\bibnamefont
  {Sheikh}}, \bibinfo {author} {\bibfnamefont {E.}~\bibnamefont
  {Juarez-Hernandez}},  \emph {et~al.},\ }\bibfield  {title} {\enquote
  {\bibinfo {title} {Cmos-based cryogenic control of silicon quantum
  circuits},}\ }\href@noop {} {\bibfield  {journal} {\bibinfo  {journal}
  {Nature}\ }\textbf {\bibinfo {volume} {593}},\ \bibinfo {pages} {205--210}
  (\bibinfo {year} {2021})}\BibitemShut {NoStop}%
\bibitem [{\citenamefont {Pauka}\ \emph {et~al.}(2021)\citenamefont {Pauka},
  \citenamefont {Das}, \citenamefont {Kalra}, \citenamefont {Moini},
  \citenamefont {Yang}, \citenamefont {Trainer}, \citenamefont {Bousquet},
  \citenamefont {Cantaloube}, \citenamefont {Dick}, \citenamefont {Gardner}
  \emph {et~al.}}]{pauka2021cryogenic}%
  \BibitemOpen
  \bibfield  {author} {\bibinfo {author} {\bibfnamefont {S.}~\bibnamefont
  {Pauka}}, \bibinfo {author} {\bibfnamefont {K.}~\bibnamefont {Das}}, \bibinfo
  {author} {\bibfnamefont {R.}~\bibnamefont {Kalra}}, \bibinfo {author}
  {\bibfnamefont {A.}~\bibnamefont {Moini}}, \bibinfo {author} {\bibfnamefont
  {Y.}~\bibnamefont {Yang}}, \bibinfo {author} {\bibfnamefont {M.}~\bibnamefont
  {Trainer}}, \bibinfo {author} {\bibfnamefont {A.}~\bibnamefont {Bousquet}},
  \bibinfo {author} {\bibfnamefont {C.}~\bibnamefont {Cantaloube}}, \bibinfo
  {author} {\bibfnamefont {N.}~\bibnamefont {Dick}}, \bibinfo {author}
  {\bibfnamefont {G.}~\bibnamefont {Gardner}},  \emph {et~al.},\ }\bibfield
  {title} {\enquote {\bibinfo {title} {A cryogenic cmos chip for generating
  control signals for multiple qubits},}\ }\href@noop {} {\bibfield  {journal}
  {\bibinfo  {journal} {Nature Electronics}\ }\textbf {\bibinfo {volume} {4}},\
  \bibinfo {pages} {64--70} (\bibinfo {year} {2021})}\BibitemShut {NoStop}%
\bibitem [{\citenamefont {Bartee}\ \emph {et~al.}(2024)\citenamefont {Bartee},
  \citenamefont {Gilbert}, \citenamefont {Zuo}, \citenamefont {Das},
  \citenamefont {Tanttu}, \citenamefont {Yang}, \citenamefont {Stuyck},
  \citenamefont {Pauka}, \citenamefont {Su}, \citenamefont {Lim} \emph
  {et~al.}}]{bartee2024spin}%
  \BibitemOpen
  \bibfield  {author} {\bibinfo {author} {\bibfnamefont {S.~K.}\ \bibnamefont
  {Bartee}}, \bibinfo {author} {\bibfnamefont {W.}~\bibnamefont {Gilbert}},
  \bibinfo {author} {\bibfnamefont {K.}~\bibnamefont {Zuo}}, \bibinfo {author}
  {\bibfnamefont {K.}~\bibnamefont {Das}}, \bibinfo {author} {\bibfnamefont
  {T.}~\bibnamefont {Tanttu}}, \bibinfo {author} {\bibfnamefont {C.~H.}\
  \bibnamefont {Yang}}, \bibinfo {author} {\bibfnamefont {N.~D.}\ \bibnamefont
  {Stuyck}}, \bibinfo {author} {\bibfnamefont {S.~J.}\ \bibnamefont {Pauka}},
  \bibinfo {author} {\bibfnamefont {R.~Y.}\ \bibnamefont {Su}}, \bibinfo
  {author} {\bibfnamefont {W.~H.}\ \bibnamefont {Lim}},  \emph {et~al.},\
  }\bibfield  {title} {\enquote {\bibinfo {title} {Spin qubits with scalable
  milli-kelvin cmos control},}\ }\href@noop {} {\bibfield  {journal} {\bibinfo
  {journal} {arXiv preprint arXiv:2407.15151}\ } (\bibinfo {year}
  {2024})}\BibitemShut {NoStop}%
\bibitem [{\citenamefont {Petit}\ \emph {et~al.}(2018)\citenamefont {Petit},
  \citenamefont {Boter}, \citenamefont {Eenink}, \citenamefont {Droulers},
  \citenamefont {Tagliaferri}, \citenamefont {Li}, \citenamefont {Franke},
  \citenamefont {Singh}, \citenamefont {Clarke}, \citenamefont {Schouten},
  \citenamefont {Dobrovitski}, \citenamefont {Vandersypen},\ and\ \citenamefont
  {Veldhorst}}]{Petit2018}%
  \BibitemOpen
  \bibfield  {author} {\bibinfo {author} {\bibfnamefont {L.}~\bibnamefont
  {Petit}}, \bibinfo {author} {\bibfnamefont {J.~M.}\ \bibnamefont {Boter}},
  \bibinfo {author} {\bibfnamefont {H.~G.~J.}\ \bibnamefont {Eenink}}, \bibinfo
  {author} {\bibfnamefont {G.}~\bibnamefont {Droulers}}, \bibinfo {author}
  {\bibfnamefont {M.~L.~V.}\ \bibnamefont {Tagliaferri}}, \bibinfo {author}
  {\bibfnamefont {R.}~\bibnamefont {Li}}, \bibinfo {author} {\bibfnamefont
  {D.~P.}\ \bibnamefont {Franke}}, \bibinfo {author} {\bibfnamefont {K.~J.}\
  \bibnamefont {Singh}}, \bibinfo {author} {\bibfnamefont {J.~S.}\ \bibnamefont
  {Clarke}}, \bibinfo {author} {\bibfnamefont {R.~N.}\ \bibnamefont
  {Schouten}}, \bibinfo {author} {\bibfnamefont {V.~V.}\ \bibnamefont
  {Dobrovitski}}, \bibinfo {author} {\bibfnamefont {L.~M.~K.}\ \bibnamefont
  {Vandersypen}}, \ and\ \bibinfo {author} {\bibfnamefont {M.}~\bibnamefont
  {Veldhorst}},\ }\bibfield  {title} {\enquote {\bibinfo {title} {Spin lifetime
  and charge noise in hot silicon quantum dot qubits},}\ }\href {\doibase
  10.1103/PhysRevLett.121.076801} {\bibfield  {journal} {\bibinfo  {journal}
  {Phys. Rev. Lett.}\ }\textbf {\bibinfo {volume} {121}},\ \bibinfo {pages}
  {076801} (\bibinfo {year} {2018})}\BibitemShut {NoStop}%
\bibitem [{\citenamefont {Ono}, \citenamefont {Mori},\ and\ \citenamefont
  {Moriyama}(2019)}]{Ono2019}%
  \BibitemOpen
  \bibfield  {author} {\bibinfo {author} {\bibfnamefont {K.}~\bibnamefont
  {Ono}}, \bibinfo {author} {\bibfnamefont {T.}~\bibnamefont {Mori}}, \ and\
  \bibinfo {author} {\bibfnamefont {S.}~\bibnamefont {Moriyama}},\ }\bibfield
  {title} {\enquote {\bibinfo {title} {High-temperature operation of a silicon
  qubit},}\ }\href@noop {} {\bibfield  {journal} {\bibinfo  {journal}
  {Scientific reports}\ }\textbf {\bibinfo {volume} {9}},\ \bibinfo {pages}
  {469} (\bibinfo {year} {2019})}\BibitemShut {NoStop}%
\bibitem [{\citenamefont {Petit}\ \emph {et~al.}(2020)\citenamefont {Petit},
  \citenamefont {Eenink}, \citenamefont {Russ}, \citenamefont {Lawrie},
  \citenamefont {Hendrickx}, \citenamefont {Philips}, \citenamefont {Clarke},
  \citenamefont {Vandersypen},\ and\ \citenamefont {Veldhorst}}]{Petit2020}%
  \BibitemOpen
  \bibfield  {author} {\bibinfo {author} {\bibfnamefont {L.}~\bibnamefont
  {Petit}}, \bibinfo {author} {\bibfnamefont {H.~G.~J.}\ \bibnamefont
  {Eenink}}, \bibinfo {author} {\bibfnamefont {M.}~\bibnamefont {Russ}},
  \bibinfo {author} {\bibfnamefont {W.~I.~L.}\ \bibnamefont {Lawrie}}, \bibinfo
  {author} {\bibfnamefont {N.~W.}\ \bibnamefont {Hendrickx}}, \bibinfo {author}
  {\bibfnamefont {S.~G.~J.}\ \bibnamefont {Philips}}, \bibinfo {author}
  {\bibfnamefont {J.~S.}\ \bibnamefont {Clarke}}, \bibinfo {author}
  {\bibfnamefont {L.~M.~K.}\ \bibnamefont {Vandersypen}}, \ and\ \bibinfo
  {author} {\bibfnamefont {M.}~\bibnamefont {Veldhorst}},\ }\bibfield  {title}
  {\enquote {\bibinfo {title} {Universal quantum logic in hot silicon
  qubits},}\ }\href@noop {} {\bibfield  {journal} {\bibinfo  {journal}
  {Nature}\ }\textbf {\bibinfo {volume} {580}},\ \bibinfo {pages} {355--359}
  (\bibinfo {year} {2020})}\BibitemShut {NoStop}%
\bibitem [{\citenamefont {Yang}\ \emph {et~al.}(2020)\citenamefont {Yang},
  \citenamefont {Leon}, \citenamefont {Hwang}, \citenamefont {Saraiva},
  \citenamefont {Tanttu}, \citenamefont {Huang}, \citenamefont {Lemyre},
  \citenamefont {Chan}, \citenamefont {Tan}, \citenamefont {Hudson} \emph
  {et~al.}}]{Yang2020}%
  \BibitemOpen
  \bibfield  {author} {\bibinfo {author} {\bibfnamefont {C.~H.}\ \bibnamefont
  {Yang}}, \bibinfo {author} {\bibfnamefont {R.~C.~C.}\ \bibnamefont {Leon}},
  \bibinfo {author} {\bibfnamefont {J.~C.~C.}\ \bibnamefont {Hwang}}, \bibinfo
  {author} {\bibfnamefont {A.}~\bibnamefont {Saraiva}}, \bibinfo {author}
  {\bibfnamefont {T.}~\bibnamefont {Tanttu}}, \bibinfo {author} {\bibfnamefont
  {W.}~\bibnamefont {Huang}}, \bibinfo {author} {\bibfnamefont {J.~C.}\
  \bibnamefont {Lemyre}}, \bibinfo {author} {\bibfnamefont {K.~W.}\
  \bibnamefont {Chan}}, \bibinfo {author} {\bibfnamefont {K.~Y.}\ \bibnamefont
  {Tan}}, \bibinfo {author} {\bibfnamefont {F.~E.}\ \bibnamefont {Hudson}},
  \emph {et~al.},\ }\bibfield  {title} {\enquote {\bibinfo {title} {Operation
  of a silicon quantum processor unit cell above one kelvin},}\ }\href@noop {}
  {\bibfield  {journal} {\bibinfo  {journal} {Nature}\ }\textbf {\bibinfo
  {volume} {580}},\ \bibinfo {pages} {350--354} (\bibinfo {year}
  {2020})}\BibitemShut {NoStop}%
\bibitem [{\citenamefont {Petit}\ \emph {et~al.}(2022)\citenamefont {Petit},
  \citenamefont {Russ}, \citenamefont {Eenink}, \citenamefont {Lawrie},
  \citenamefont {Clarke}, \citenamefont {Vandersypen},\ and\ \citenamefont
  {Veldhorst}}]{Petit2022}%
  \BibitemOpen
  \bibfield  {author} {\bibinfo {author} {\bibfnamefont {L.}~\bibnamefont
  {Petit}}, \bibinfo {author} {\bibfnamefont {M.}~\bibnamefont {Russ}},
  \bibinfo {author} {\bibfnamefont {G.~H. G.~J.}\ \bibnamefont {Eenink}},
  \bibinfo {author} {\bibfnamefont {W.~I.~L.}\ \bibnamefont {Lawrie}}, \bibinfo
  {author} {\bibfnamefont {J.~S.}\ \bibnamefont {Clarke}}, \bibinfo {author}
  {\bibfnamefont {L.~M.~K.}\ \bibnamefont {Vandersypen}}, \ and\ \bibinfo
  {author} {\bibfnamefont {M.}~\bibnamefont {Veldhorst}},\ }\bibfield  {title}
  {\enquote {\bibinfo {title} {Design and integration of single-qubit rotations
  and two-qubit gates in silicon above one kelvin},}\ }\href@noop {} {\bibfield
   {journal} {\bibinfo  {journal} {Communications Materials}\ }\textbf
  {\bibinfo {volume} {3}},\ \bibinfo {pages} {82} (\bibinfo {year}
  {2022})}\BibitemShut {NoStop}%
\bibitem [{\citenamefont {Camenzind}\ \emph {et~al.}(2022)\citenamefont
  {Camenzind}, \citenamefont {Geyer}, \citenamefont {Fuhrer}, \citenamefont
  {Warburton}, \citenamefont {Zumb{\"u}hl},\ and\ \citenamefont
  {Kuhlmann}}]{Camenzind2022}%
  \BibitemOpen
  \bibfield  {author} {\bibinfo {author} {\bibfnamefont {L.~C.}\ \bibnamefont
  {Camenzind}}, \bibinfo {author} {\bibfnamefont {S.}~\bibnamefont {Geyer}},
  \bibinfo {author} {\bibfnamefont {A.}~\bibnamefont {Fuhrer}}, \bibinfo
  {author} {\bibfnamefont {R.~J.}\ \bibnamefont {Warburton}}, \bibinfo {author}
  {\bibfnamefont {D.~M.}\ \bibnamefont {Zumb{\"u}hl}}, \ and\ \bibinfo {author}
  {\bibfnamefont {A.~V.}\ \bibnamefont {Kuhlmann}},\ }\bibfield  {title}
  {\enquote {\bibinfo {title} {A hole spin qubit in a fin field-effect
  transistor above 4 kelvin},}\ }\href@noop {} {\bibfield  {journal} {\bibinfo
  {journal} {Nature Electronics}\ }\textbf {\bibinfo {volume} {5}},\ \bibinfo
  {pages} {178--183} (\bibinfo {year} {2022})}\BibitemShut {NoStop}%
\bibitem [{\citenamefont {Beenakker}(1991)}]{Beenakker1991}%
  \BibitemOpen
  \bibfield  {author} {\bibinfo {author} {\bibfnamefont {C.~W.~J.}\
  \bibnamefont {Beenakker}},\ }\bibfield  {title} {\enquote {\bibinfo {title}
  {Theory of coulomb-blockade oscillations in the conductance of a quantum
  dot},}\ }\href {\doibase 10.1103/PhysRevB.44.1646} {\bibfield  {journal}
  {\bibinfo  {journal} {Phys. Rev. B}\ }\textbf {\bibinfo {volume} {44}},\
  \bibinfo {pages} {1646--1656} (\bibinfo {year} {1991})}\BibitemShut {NoStop}%
\bibitem [{\citenamefont {Giazotto}\ \emph {et~al.}(2006)\citenamefont
  {Giazotto}, \citenamefont {Heikkil{\"a}}, \citenamefont {Luukanen},
  \citenamefont {Savin},\ and\ \citenamefont {Pekola}}]{Giazotto2006}%
  \BibitemOpen
  \bibfield  {author} {\bibinfo {author} {\bibfnamefont {F.}~\bibnamefont
  {Giazotto}}, \bibinfo {author} {\bibfnamefont {T.~T.}\ \bibnamefont
  {Heikkil{\"a}}}, \bibinfo {author} {\bibfnamefont {A.}~\bibnamefont
  {Luukanen}}, \bibinfo {author} {\bibfnamefont {A.~M.}\ \bibnamefont {Savin}},
  \ and\ \bibinfo {author} {\bibfnamefont {J.~P.}\ \bibnamefont {Pekola}},\
  }\bibfield  {title} {\enquote {\bibinfo {title} {Opportunities for
  mesoscopics in thermometry and refrigeration: Physics and applications},}\
  }\href@noop {} {\bibfield  {journal} {\bibinfo  {journal} {Reviews of Modern
  Physics}\ }\textbf {\bibinfo {volume} {78}},\ \bibinfo {pages} {217}
  (\bibinfo {year} {2006})}\BibitemShut {NoStop}%
\bibitem [{\citenamefont {Rossi}, \citenamefont {Ferrus},\ and\ \citenamefont
  {Williams}(2012)}]{rossi2012electron}%
  \BibitemOpen
  \bibfield  {author} {\bibinfo {author} {\bibfnamefont {A.}~\bibnamefont
  {Rossi}}, \bibinfo {author} {\bibfnamefont {T.}~\bibnamefont {Ferrus}}, \
  and\ \bibinfo {author} {\bibfnamefont {D.}~\bibnamefont {Williams}},\
  }\bibfield  {title} {\enquote {\bibinfo {title} {Electron temperature in
  electrically isolated si double quantum dots},}\ }\href@noop {} {\bibfield
  {journal} {\bibinfo  {journal} {Applied Physics Letters}\ }\textbf {\bibinfo
  {volume} {100}} (\bibinfo {year} {2012})}\BibitemShut {NoStop}%
\bibitem [{\citenamefont {Molenkamp}\ \emph {et~al.}(1992)\citenamefont
  {Molenkamp}, \citenamefont {Gravier}, \citenamefont {van Houten},
  \citenamefont {Buijk}, \citenamefont {Mabesoone},\ and\ \citenamefont
  {Foxon}}]{Molenkamp1992}%
  \BibitemOpen
  \bibfield  {author} {\bibinfo {author} {\bibfnamefont {L.~W.}\ \bibnamefont
  {Molenkamp}}, \bibinfo {author} {\bibfnamefont {T.}~\bibnamefont {Gravier}},
  \bibinfo {author} {\bibfnamefont {H.}~\bibnamefont {van Houten}}, \bibinfo
  {author} {\bibfnamefont {O.~J.~A.}\ \bibnamefont {Buijk}}, \bibinfo {author}
  {\bibfnamefont {M.~A.~A.}\ \bibnamefont {Mabesoone}}, \ and\ \bibinfo
  {author} {\bibfnamefont {C.~T.}\ \bibnamefont {Foxon}},\ }\bibfield  {title}
  {\enquote {\bibinfo {title} {Peltier coefficient and thermal conductance of a
  quantum point contact},}\ }\href {\doibase 10.1103/PhysRevLett.68.3765}
  {\bibfield  {journal} {\bibinfo  {journal} {Phys. Rev. Lett.}\ }\textbf
  {\bibinfo {volume} {68}},\ \bibinfo {pages} {3765--3768} (\bibinfo {year}
  {1992})}\BibitemShut {NoStop}%
\bibitem [{\citenamefont {Chiatti}\ \emph {et~al.}(2006)\citenamefont
  {Chiatti}, \citenamefont {Nicholls}, \citenamefont {Proskuryakov},
  \citenamefont {Lumpkin}, \citenamefont {Farrer},\ and\ \citenamefont
  {Ritchie}}]{Chiatti2006}%
  \BibitemOpen
  \bibfield  {author} {\bibinfo {author} {\bibfnamefont {O.}~\bibnamefont
  {Chiatti}}, \bibinfo {author} {\bibfnamefont {J.~T.}\ \bibnamefont
  {Nicholls}}, \bibinfo {author} {\bibfnamefont {Y.~Y.}\ \bibnamefont
  {Proskuryakov}}, \bibinfo {author} {\bibfnamefont {N.}~\bibnamefont
  {Lumpkin}}, \bibinfo {author} {\bibfnamefont {I.}~\bibnamefont {Farrer}}, \
  and\ \bibinfo {author} {\bibfnamefont {D.~A.}\ \bibnamefont {Ritchie}},\
  }\bibfield  {title} {\enquote {\bibinfo {title} {Quantum thermal conductance
  of electrons in a one-dimensional wire},}\ }\href {\doibase
  10.1103/PhysRevLett.97.056601} {\bibfield  {journal} {\bibinfo  {journal}
  {Phys. Rev. Lett.}\ }\textbf {\bibinfo {volume} {97}},\ \bibinfo {pages}
  {056601} (\bibinfo {year} {2006})}\BibitemShut {NoStop}%
\bibitem [{\citenamefont {Hoffmann}\ \emph {et~al.}(2007)\citenamefont
  {Hoffmann}, \citenamefont {Nakpathomkun}, \citenamefont {Persson},
  \citenamefont {Linke}, \citenamefont {Nilsson},\ and\ \citenamefont
  {Samuelson}}]{Hoffmann2007}%
  \BibitemOpen
  \bibfield  {author} {\bibinfo {author} {\bibfnamefont {E.~A.}\ \bibnamefont
  {Hoffmann}}, \bibinfo {author} {\bibfnamefont {N.}~\bibnamefont
  {Nakpathomkun}}, \bibinfo {author} {\bibfnamefont {A.~I.}\ \bibnamefont
  {Persson}}, \bibinfo {author} {\bibfnamefont {H.}~\bibnamefont {Linke}},
  \bibinfo {author} {\bibfnamefont {H.~A.}\ \bibnamefont {Nilsson}}, \ and\
  \bibinfo {author} {\bibfnamefont {L.}~\bibnamefont {Samuelson}},\ }\bibfield
  {title} {\enquote {\bibinfo {title} {Quantum-dot thermometry},}\ }\href@noop
  {} {\bibfield  {journal} {\bibinfo  {journal} {Applied Physics Letters}\
  }\textbf {\bibinfo {volume} {91}} (\bibinfo {year} {2007})}\BibitemShut
  {NoStop}%
\bibitem [{\citenamefont {Hoffmann}\ \emph {et~al.}(2009)\citenamefont
  {Hoffmann}, \citenamefont {Nilsson}, \citenamefont {Matthews}, \citenamefont
  {Nakpathomkun}, \citenamefont {Persson}, \citenamefont {Samuelson},\ and\
  \citenamefont {Linke}}]{Hoffmann2009}%
  \BibitemOpen
  \bibfield  {author} {\bibinfo {author} {\bibfnamefont {E.~A.}\ \bibnamefont
  {Hoffmann}}, \bibinfo {author} {\bibfnamefont {H.~A.}\ \bibnamefont
  {Nilsson}}, \bibinfo {author} {\bibfnamefont {J.~E.}\ \bibnamefont
  {Matthews}}, \bibinfo {author} {\bibfnamefont {N.}~\bibnamefont
  {Nakpathomkun}}, \bibinfo {author} {\bibfnamefont {A.~I.}\ \bibnamefont
  {Persson}}, \bibinfo {author} {\bibfnamefont {L.}~\bibnamefont {Samuelson}},
  \ and\ \bibinfo {author} {\bibfnamefont {H.}~\bibnamefont {Linke}},\
  }\bibfield  {title} {\enquote {\bibinfo {title} {Measuring temperature
  gradients over nanometer length scales},}\ }\href@noop {} {\bibfield
  {journal} {\bibinfo  {journal} {Nano Letters}\ }\textbf {\bibinfo {volume}
  {9}},\ \bibinfo {pages} {779--783} (\bibinfo {year} {2009})}\BibitemShut
  {NoStop}%
\bibitem [{\citenamefont {Dutta}\ \emph {et~al.}(2017)\citenamefont {Dutta},
  \citenamefont {Peltonen}, \citenamefont {Antonenko}, \citenamefont {Meschke},
  \citenamefont {Skvortsov}, \citenamefont {Kubala}, \citenamefont {K\"onig},
  \citenamefont {Winkelmann}, \citenamefont {Courtois},\ and\ \citenamefont
  {Pekola}}]{Dutta2017}%
  \BibitemOpen
  \bibfield  {author} {\bibinfo {author} {\bibfnamefont {B.}~\bibnamefont
  {Dutta}}, \bibinfo {author} {\bibfnamefont {J.~T.}\ \bibnamefont {Peltonen}},
  \bibinfo {author} {\bibfnamefont {D.~S.}\ \bibnamefont {Antonenko}}, \bibinfo
  {author} {\bibfnamefont {M.}~\bibnamefont {Meschke}}, \bibinfo {author}
  {\bibfnamefont {M.~A.}\ \bibnamefont {Skvortsov}}, \bibinfo {author}
  {\bibfnamefont {B.}~\bibnamefont {Kubala}}, \bibinfo {author} {\bibfnamefont
  {J.}~\bibnamefont {K\"onig}}, \bibinfo {author} {\bibfnamefont {C.~B.}\
  \bibnamefont {Winkelmann}}, \bibinfo {author} {\bibfnamefont
  {H.}~\bibnamefont {Courtois}}, \ and\ \bibinfo {author} {\bibfnamefont
  {J.~P.}\ \bibnamefont {Pekola}},\ }\bibfield  {title} {\enquote {\bibinfo
  {title} {Thermal conductance of a single-electron transistor},}\ }\href
  {\doibase 10.1103/PhysRevLett.119.077701} {\bibfield  {journal} {\bibinfo
  {journal} {Phys. Rev. Lett.}\ }\textbf {\bibinfo {volume} {119}},\ \bibinfo
  {pages} {077701} (\bibinfo {year} {2017})}\BibitemShut {NoStop}%
\bibitem [{\citenamefont {de~Kruijf}\ \emph {et~al.}(2024)\citenamefont
  {de~Kruijf}, \citenamefont {Noah}, \citenamefont {Gomez-Saiz}, \citenamefont
  {Morton},\ and\ \citenamefont {Gonzalez-Zalba}}]{de2023measurement}%
  \BibitemOpen
  \bibfield  {author} {\bibinfo {author} {\bibfnamefont {M.}~\bibnamefont
  {de~Kruijf}}, \bibinfo {author} {\bibfnamefont {G.~M.}\ \bibnamefont {Noah}},
  \bibinfo {author} {\bibfnamefont {A.}~\bibnamefont {Gomez-Saiz}}, \bibinfo
  {author} {\bibfnamefont {J.~J.}\ \bibnamefont {Morton}}, \ and\ \bibinfo
  {author} {\bibfnamefont {M.~F.}\ \bibnamefont {Gonzalez-Zalba}},\ }\bibfield
  {title} {\enquote {\bibinfo {title} {Measurement of cryoelectronics heating
  using a local quantum dot thermometer in silicon},}\ }\href@noop {}
  {\bibfield  {journal} {\bibinfo  {journal} {Chip}\ ,\ \bibinfo {pages}
  {100097}} (\bibinfo {year} {2024})}\BibitemShut {NoStop}%
\bibitem [{\citenamefont {Champain}\ \emph {et~al.}(2024)\citenamefont
  {Champain}, \citenamefont {Schmitt}, \citenamefont {Bertrand}, \citenamefont
  {Niebojewski}, \citenamefont {Maurand}, \citenamefont {Jehl}, \citenamefont
  {Winkelmann}, \citenamefont {De~Franceschi},\ and\ \citenamefont
  {Brun}}]{champain2023real}%
  \BibitemOpen
  \bibfield  {author} {\bibinfo {author} {\bibfnamefont {V.}~\bibnamefont
  {Champain}}, \bibinfo {author} {\bibfnamefont {V.}~\bibnamefont {Schmitt}},
  \bibinfo {author} {\bibfnamefont {B.}~\bibnamefont {Bertrand}}, \bibinfo
  {author} {\bibfnamefont {H.}~\bibnamefont {Niebojewski}}, \bibinfo {author}
  {\bibfnamefont {R.}~\bibnamefont {Maurand}}, \bibinfo {author} {\bibfnamefont
  {X.}~\bibnamefont {Jehl}}, \bibinfo {author} {\bibfnamefont {C.}~\bibnamefont
  {Winkelmann}}, \bibinfo {author} {\bibfnamefont {S.}~\bibnamefont
  {De~Franceschi}}, \ and\ \bibinfo {author} {\bibfnamefont {B.}~\bibnamefont
  {Brun}},\ }\bibfield  {title} {\enquote {\bibinfo {title} {Real-time
  millikelvin thermometry in a semiconductor-qubit architecture},}\ }\href
  {\doibase 10.1103/PhysRevApplied.21.064039} {\bibfield  {journal} {\bibinfo
  {journal} {Phys. Rev. Appl.}\ }\textbf {\bibinfo {volume} {21}},\ \bibinfo
  {pages} {064039} (\bibinfo {year} {2024})}\BibitemShut {NoStop}%
\bibitem [{\citenamefont {Duthil}(2014)}]{metal}%
  \BibitemOpen
  \bibfield  {author} {\bibinfo {author} {\bibfnamefont {P.}~\bibnamefont
  {Duthil}},\ }\bibfield  {title} {\enquote {\bibinfo {title} {Material
  properties at low temperature},}\ }\href@noop {} {\bibfield  {journal}
  {\bibinfo  {journal} {CAS - CERN Accelerator School}\ ,\ \bibinfo {pages}
  {77–95}} (\bibinfo {year} {2014})}\BibitemShut {NoStop}%
\bibitem [{\citenamefont {Ekin}(2006)}]{ekin}%
  \BibitemOpen
  \bibfield  {author} {\bibinfo {author} {\bibfnamefont {J.}~\bibnamefont
  {Ekin}},\ }\href@noop {} {\emph {\bibinfo {title} {Experimental techniques
  for low-temperature measurements: cryostat design, material properties and
  superconductor critical-current testing}}}\ (\bibinfo  {publisher} {Oxford
  university press},\ \bibinfo {year} {2006})\BibitemShut {NoStop}%
\bibitem [{ais()}]{aist}%
  \BibitemOpen
  \href {https://tpds.db.aist.go.jp/index_en.html} {\emph {\bibinfo {title}
  {Thermophysical Property Database}}},\ \bibinfo {organization} {National
  Institute of Advanced Industrial Science and Technology}\BibitemShut
  {NoStop}%
\bibitem [{\citenamefont {Sato}\ and\ \citenamefont
  {Kawahara}(2024)}]{sato2024simulation}%
  \BibitemOpen
  \bibfield  {author} {\bibinfo {author} {\bibfnamefont {Y.}~\bibnamefont
  {Sato}}\ and\ \bibinfo {author} {\bibfnamefont {T.}~\bibnamefont
  {Kawahara}},\ }\bibfield  {title} {\enquote {\bibinfo {title} {Simulation of
  temperature-dependent quantum gates in silicon quantum dots with frequency
  shifts},}\ }\href@noop {} {\bibfield  {journal} {\bibinfo  {journal} {arXiv
  preprint arXiv:2407.05295}\ } (\bibinfo {year} {2024})}\BibitemShut {NoStop}%
\bibitem [{\citenamefont {Utsugi}\ \emph {et~al.}(2023)\citenamefont {Utsugi},
  \citenamefont {Lee}, \citenamefont {Tsuchiya}, \citenamefont {Mine},
  \citenamefont {Mizokuchi}, \citenamefont {Yoneda}, \citenamefont {Kodera},
  \citenamefont {Saito}, \citenamefont {Hisamoto},\ and\ \citenamefont
  {Mizuno}}]{utsugi2023single}%
  \BibitemOpen
  \bibfield  {author} {\bibinfo {author} {\bibfnamefont {T.}~\bibnamefont
  {Utsugi}}, \bibinfo {author} {\bibfnamefont {N.}~\bibnamefont {Lee}},
  \bibinfo {author} {\bibfnamefont {R.}~\bibnamefont {Tsuchiya}}, \bibinfo
  {author} {\bibfnamefont {T.}~\bibnamefont {Mine}}, \bibinfo {author}
  {\bibfnamefont {R.}~\bibnamefont {Mizokuchi}}, \bibinfo {author}
  {\bibfnamefont {J.}~\bibnamefont {Yoneda}}, \bibinfo {author} {\bibfnamefont
  {T.}~\bibnamefont {Kodera}}, \bibinfo {author} {\bibfnamefont
  {S.}~\bibnamefont {Saito}}, \bibinfo {author} {\bibfnamefont
  {D.}~\bibnamefont {Hisamoto}}, \ and\ \bibinfo {author} {\bibfnamefont
  {H.}~\bibnamefont {Mizuno}},\ }\bibfield  {title} {\enquote {\bibinfo {title}
  {Single-electron pump in a quantum dot array for silicon quantum
  computers},}\ }\href@noop {} {\bibfield  {journal} {\bibinfo  {journal}
  {Japanese Journal of Applied Physics}\ }\textbf {\bibinfo {volume} {62}},\
  \bibinfo {pages} {SC1020} (\bibinfo {year} {2023})}\BibitemShut {NoStop}%
\bibitem [{\citenamefont {Kuno}\ \emph {et~al.}(2025)\citenamefont {Kuno},
  \citenamefont {Utsugi}, \citenamefont {Tsuchiya}, \citenamefont {Lee},
  \citenamefont {Mine}, \citenamefont {Yanagi}, \citenamefont {Mizokuchi},
  \citenamefont {Yoneda}, \citenamefont {Kodera}, \citenamefont {Saito} \emph
  {et~al.}}]{kuno2024}%
  \BibitemOpen
  \bibfield  {author} {\bibinfo {author} {\bibfnamefont {T.}~\bibnamefont
  {Kuno}}, \bibinfo {author} {\bibfnamefont {T.}~\bibnamefont {Utsugi}},
  \bibinfo {author} {\bibfnamefont {R.}~\bibnamefont {Tsuchiya}}, \bibinfo
  {author} {\bibfnamefont {N.}~\bibnamefont {Lee}}, \bibinfo {author}
  {\bibfnamefont {T.}~\bibnamefont {Mine}}, \bibinfo {author} {\bibfnamefont
  {I.}~\bibnamefont {Yanagi}}, \bibinfo {author} {\bibfnamefont
  {R.}~\bibnamefont {Mizokuchi}}, \bibinfo {author} {\bibfnamefont
  {J.}~\bibnamefont {Yoneda}}, \bibinfo {author} {\bibfnamefont
  {T.}~\bibnamefont {Kodera}}, \bibinfo {author} {\bibfnamefont
  {S.}~\bibnamefont {Saito}},  \emph {et~al.},\ }\bibfield  {title} {\enquote
  {\bibinfo {title} {Single-electron charge sensor self-aligned to a quantum
  dot array by double-gate patterning process},}\ }\href@noop {} {\bibfield
  {journal} {\bibinfo  {journal} {Japanese Journal of Applied Physics}\
  }\textbf {\bibinfo {volume} {64}},\ \bibinfo {pages} {011001} (\bibinfo
  {year} {2025})}\BibitemShut {NoStop}%
\bibitem [{\citenamefont {Kouwenhoven}\ \emph {et~al.}(1997)\citenamefont
  {Kouwenhoven}, \citenamefont {Marcus}, \citenamefont {McEuen}, \citenamefont
  {Tarucha}, \citenamefont {Westervelt},\ and\ \citenamefont
  {Wingreen}}]{kouwenhoven1997electron}%
  \BibitemOpen
  \bibfield  {author} {\bibinfo {author} {\bibfnamefont {L.~P.}\ \bibnamefont
  {Kouwenhoven}}, \bibinfo {author} {\bibfnamefont {C.~M.}\ \bibnamefont
  {Marcus}}, \bibinfo {author} {\bibfnamefont {P.~L.}\ \bibnamefont {McEuen}},
  \bibinfo {author} {\bibfnamefont {S.}~\bibnamefont {Tarucha}}, \bibinfo
  {author} {\bibfnamefont {R.~M.}\ \bibnamefont {Westervelt}}, \ and\ \bibinfo
  {author} {\bibfnamefont {N.~S.}\ \bibnamefont {Wingreen}},\ }\bibfield
  {title} {\enquote {\bibinfo {title} {Electron transport in quantum dots},}\
  }in\ \href@noop {} {\emph {\bibinfo {booktitle} {Mesoscopic electron
  transport}}}\ (\bibinfo  {publisher} {Springer},\ \bibinfo {year} {1997})\
  pp.\ \bibinfo {pages} {105--214}\BibitemShut {NoStop}%
\bibitem [{\citenamefont {Taguchi}\ \emph {et~al.}(2024)\citenamefont
  {Taguchi}, \citenamefont {Okidono}, \citenamefont {Miki},\ and\ \citenamefont
  {Nagata}}]{taguchi2024si}%
  \BibitemOpen
  \bibfield  {author} {\bibinfo {author} {\bibfnamefont {M.}~\bibnamefont
  {Taguchi}}, \bibinfo {author} {\bibfnamefont {T.}~\bibnamefont {Okidono}},
  \bibinfo {author} {\bibfnamefont {T.}~\bibnamefont {Miki}}, \ and\ \bibinfo
  {author} {\bibfnamefont {M.}~\bibnamefont {Nagata}},\ }\bibfield  {title}
  {\enquote {\bibinfo {title} {Si interposer with cu tsvs on cu substrate
  thermally and electrically anchoring qubit chips in millikelvin assembly},}\
  }in\ \href@noop {} {\emph {\bibinfo {booktitle} {2024 IEEE 74th Electronic
  Components and Technology Conference (ECTC)}}}\ (\bibinfo {organization}
  {IEEE},\ \bibinfo {year} {2024})\ pp.\ \bibinfo {pages}
  {447--450}\BibitemShut {NoStop}%
\bibitem [{\citenamefont {Keith}\ \emph {et~al.}(2019)\citenamefont {Keith},
  \citenamefont {Gorman}, \citenamefont {Kranz}, \citenamefont {He},
  \citenamefont {Keizer}, \citenamefont {Broome},\ and\ \citenamefont
  {Simmons}}]{keith}%
  \BibitemOpen
  \bibfield  {author} {\bibinfo {author} {\bibfnamefont {D.}~\bibnamefont
  {Keith}}, \bibinfo {author} {\bibfnamefont {S.}~\bibnamefont {Gorman}},
  \bibinfo {author} {\bibfnamefont {L.}~\bibnamefont {Kranz}}, \bibinfo
  {author} {\bibfnamefont {Y.}~\bibnamefont {He}}, \bibinfo {author}
  {\bibfnamefont {J.}~\bibnamefont {Keizer}}, \bibinfo {author} {\bibfnamefont
  {M.}~\bibnamefont {Broome}}, \ and\ \bibinfo {author} {\bibfnamefont
  {M.}~\bibnamefont {Simmons}},\ }\bibfield  {title} {\enquote {\bibinfo
  {title} {Benchmarking high fidelity single-shot readout of semiconductor
  qubits},}\ }\href@noop {} {\bibfield  {journal} {\bibinfo  {journal} {New
  Journal of Physics}\ }\textbf {\bibinfo {volume} {21}},\ \bibinfo {pages}
  {063011} (\bibinfo {year} {2019})}\BibitemShut {NoStop}%
\bibitem [{\citenamefont {Maire}\ \emph {et~al.}(2017)\citenamefont {Maire},
  \citenamefont {Anufriev}, \citenamefont {Yanagisawa}, \citenamefont
  {Ramiere}, \citenamefont {Volz},\ and\ \citenamefont
  {Nomura}}]{maire2017heat}%
  \BibitemOpen
  \bibfield  {author} {\bibinfo {author} {\bibfnamefont {J.}~\bibnamefont
  {Maire}}, \bibinfo {author} {\bibfnamefont {R.}~\bibnamefont {Anufriev}},
  \bibinfo {author} {\bibfnamefont {R.}~\bibnamefont {Yanagisawa}}, \bibinfo
  {author} {\bibfnamefont {A.}~\bibnamefont {Ramiere}}, \bibinfo {author}
  {\bibfnamefont {S.}~\bibnamefont {Volz}}, \ and\ \bibinfo {author}
  {\bibfnamefont {M.}~\bibnamefont {Nomura}},\ }\bibfield  {title} {\enquote
  {\bibinfo {title} {Heat conduction tuning by wave nature of phonons},}\
  }\href@noop {} {\bibfield  {journal} {\bibinfo  {journal} {Science advances}\
  }\textbf {\bibinfo {volume} {3}},\ \bibinfo {pages} {e1700027} (\bibinfo
  {year} {2017})}\BibitemShut {NoStop}%
\end{thebibliography}%

\end{document}